\documentclass[twocolumn,superscriptaddress,prl,aps,preprintnumbers,longbibliography]{revtex4-2}
\usepackage{amsmath,amssymb,bm,graphicx,color,gensymb,bbold,appendix,xspace}
\usepackage[colorlinks=true, citecolor={blue!80!black}, urlcolor={blue!50!black}, linkcolor = {blue!80!black}]{hyperref}
\usepackage[utf8x]{inputenc}

\usepackage{xcolor}

\newcommand{\KSeCo}{K$_2$Co(SeO$_3$)$_2$\xspace}

\newcommand{\be}{\begin{equation} }
	\newcommand{\ee}{\end{equation} }
\newcommand{\bea}{\begin{eqnarray} }
	\newcommand{\eea}{\end{eqnarray} }

\begin{document}

\title{Continuum excitations in a spin-supersolid on a triangular lattice}

\author{M.~Zhu}
\address{Laboratory for Solid State Physics, ETH Z\"{u}rich, 8093 Z\"{u}rich, Switzerland}
\author{V. Romerio}
\address{Laboratory for Solid State Physics, ETH Z\"{u}rich, 8093 Z\"{u}rich, Switzerland}
\author{N. Steiger}
\address{Laboratory for Solid State Physics, ETH Z\"{u}rich, 8093 Z\"{u}rich, Switzerland}

\author{S. D. Nabi}
\address{Laboratory for Solid State Physics, ETH Z\"{u}rich, 8093 Z\"{u}rich, Switzerland}

\author{N. Murai}
\address{J-PARC Center, Japan Atomic Energy Agency, Tokai, Ibaraki 319-1195, Japan}

\author{S. Ohira-Kawamura}	
\address{J-PARC Center, Japan Atomic Energy Agency, Tokai, Ibaraki 319-1195, Japan}

\author{K.~Yu.~Povarov}
\address{Dresden High Magnetic Field Laboratory (HLD-EMFL) and W\"urzburg-Dresden Cluster of Excellence ct.qmat, Helmholtz-Zentrum Dresden-Rossendorf, 01328 Dresden, Germany}
\author{Y. Skourski}
\address{Dresden High Magnetic Field Laboratory (HLD-EMFL) and W\"urzburg-Dresden Cluster of Excellence ct.qmat, Helmholtz-Zentrum Dresden-Rossendorf, 01328 Dresden, Germany}

\author{R. Sibille}
\address{Laboratory for Neutron Scattering and Imaging, Paul Scherrer Institute, CH-5232 Villigen, Switzerland}
\author{L. Keller}
\address{Laboratory for Neutron Scattering and Imaging, Paul Scherrer Institute, CH-5232 Villigen, Switzerland}

\author{Z. Yan}
\address{Laboratory for Solid State Physics, ETH Z\"{u}rich, 8093 Z\"{u}rich, Switzerland}
\author{S. Gvasaliya}
\address{Laboratory for Solid State Physics, ETH Z\"{u}rich, 8093 Z\"{u}rich, Switzerland}

\author{A.~Zheludev}
\email{zhelud@ethz.ch; http://www.neutron.ethz.ch/}
\address{Laboratory for Solid State Physics, ETH Z\"{u}rich, 8093 Z\"{u}rich, Switzerland}

\begin{abstract}
	Magnetic, thermodynamic, neutron diffraction and inelastic neutron scattering are used to study spin correlations in the easy-axis XXZ triangular lattice magnet \KSeCo. Despite the presence of quasi-2D ``supersolid'' magnetic order, the low-energy excitation spectrum contains no sharp modes and is instead a broad and structured multi-particle continuum. Applying a weak magnetic field drives the system into an $m=1/3$ fractional magnetization plateau phase and restores sharp spin wave modes. { To some extent, the behavior at zero field can be understood in terms of spin wave decay.  However, the presence of clear excitation minima at the $M$-points of the Brillouin zone suggest that the spinon language may provide a more adequate description, and signals a possible proximity to a Dirac spin liquid state.}

\end{abstract}

\date{\today}
\maketitle
The Ising antiferromagnet (AF) on a triangular lattice is the textbook example of geometric frustration \cite{Wannier1950, Miyashita1986}. The ubiquitous cartoon shows one spin pointing up, its neighbor pointing down to minimize exchange energy, this configuration leaving the  preferred direction for a third spin undefined. Beyond this simplistic picture, the problem is actually a very complex one. The quantum $S=1/2$ nearest-neighbor XXZ-model with easy-axis anisotropy is predicted to have a  peculiar ground state that can be viewed as a $\sqrt{3} \times \sqrt{3}$ ``spin-supersolid'' \cite{Heidarian2010,Yamamoto2014,Gao2022}, and a series of quantum phases in applied fields. The latter include a colinear ``up-up-down'' ($uud$) ``spin-solid'' state corresponding that is an $m=1/3$ magnetization plateau. Not much is known about excitations in that model, even as significant progress has recently been made in understanding its {\em easy plane} counterpart. There, despite the presence of long range order, the excitations are nothing like those predicted by semiclassical spin wave theory (SWT)  \cite{Ma2016,Kamiya2018,Itoh2017,Coldea2020}. Instead, they are dominated by bound states and continua of partially-free  fractional  excitations known as spinons \cite{Ghioldi2022}. This is taken as a fingerprint of proximate quantum spin-liquid states first hypothesized by Anderson \cite{Anderson1973,Balents2010} and later found in numerous triangular-lattice models (see, for instance, Refs.~\onlinecite{Iqbal2016,Wietek2017,Hu2019}). 
Does the {\em easy-axis} triangular AF feature similarly exotic spin dynamics?

We address this question experimentally and study the planar XXZ Ising-like antiferromagnet \KSeCo \cite{Zhong2020}. We show that in zero field and low temperatures the system has two-dimensional  magnetic order consistent with a spin-supersolid. The low-energy spin excitation spectrum, however, is entirely dominated by a broad gapless spinon-like continuum, rather than by sharp spin wave modes. Applying a very modest external magnetic field induces a quantum critical point. Beyond that the material enters an $m=1/3$ $uud$ plateau. The spectrum is drastically reconstructed. It consists of sharp gapped excitations that are perfectly reproduced by SWT. 

\begin{figure}
	\includegraphics[width=\columnwidth]{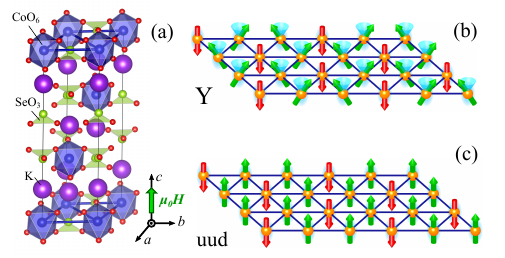}
	\caption{(a) Crystal structure of \KSeCo. (b) The ``Y'' spin-supersolid phase predicted to be the ground state of an easy-axis XXZ triangular-lattice antiferromagnet. The cyan cones symbolize rotational degeneracy. (c) A magnetic field applied along the $c$ axis stabilized a colinear ``$uud$'' spin-solid phase. }\label{struc}
\end{figure}

The first magnetic studies of \KSeCo have been reported only recently \cite{Zhong2020}. In the hexagonal structure (space group $R$-3$m$, $a=5.52$~\AA, $c=18.52$~\AA~\cite{Wildner1994}) the key features are ABC-stacked triangular planes of Co$^{2+}$ ions, as illustrated in Fig.~\ref{struc}a. As in the structurally related Na$_2$BaCo(PO$_4$)$_2$ \cite{Wellm2021,Gao2022}, the local environment is close to octahedral, ensuring a $J=\pm 1/2$ doublet single-ion ground state. The Co$^{2+}$ moments can be treated as $S=1/2$ pseudospins. The strongest magnetic interaction is between nearest-neighbor ions in each plane. Previous studies \cite{Zhong2020} have confirmed the dominance of AF coupling with very strong easy-axis type anisotropy.    In zero field the system seems to avoid any magnetic ordering down to 0.35~K. Only a broad bump is observed in specific heat at around $T\sim 1$~K, perhaps indicating short-range order. In magnetic fields exceeding 1~T directed along the $c$ axis, a sharp specific heat anomaly indicates robust long-range ordering at temperatures as high as 11~K at 9~T. It was initially suggested that this behavior corresponds to a metamagnetic transition to a fully polarized state.

\begin{figure}
 	\includegraphics[width=\columnwidth]{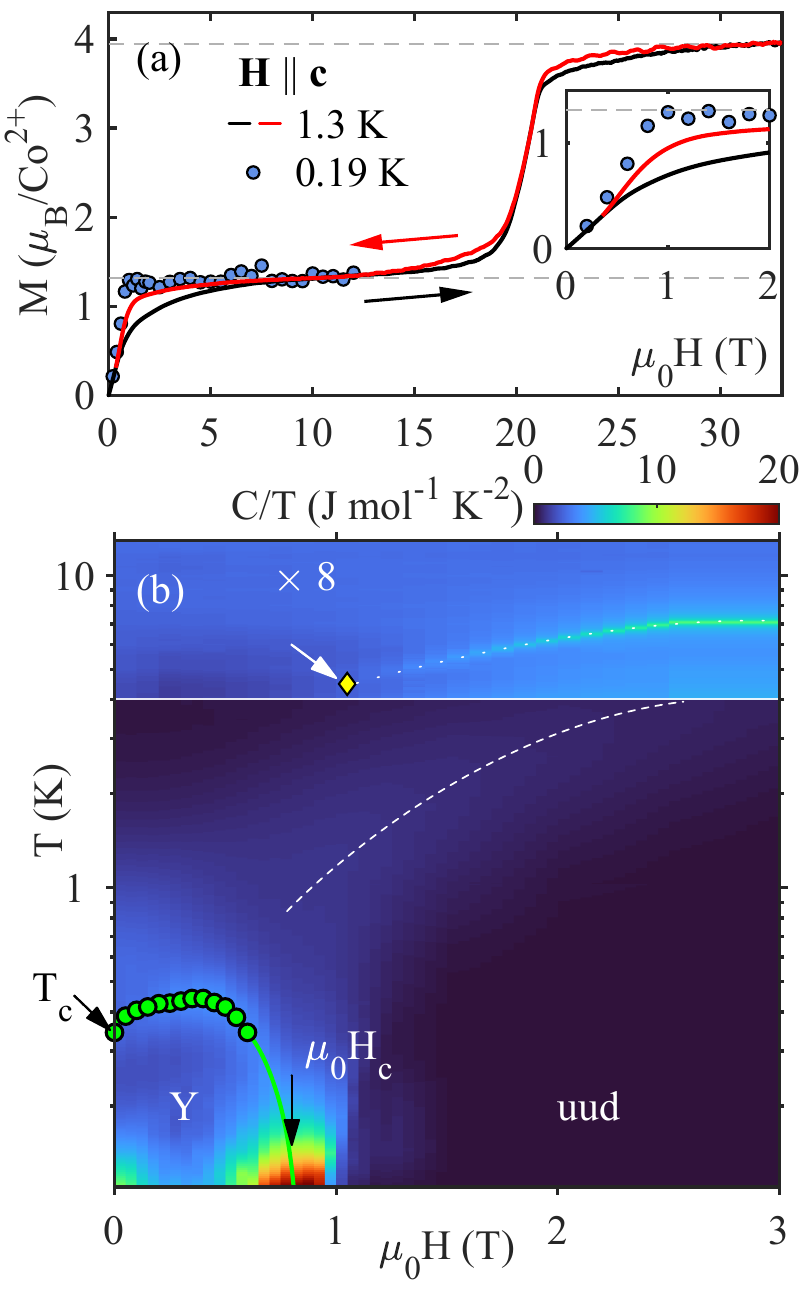}
	\caption{(a) Magnetization measured in \KSeCo in a pulsed magnetic field along the $c$ axis at $T=1.3$~K. Symbols: Faraday balance magnetometry at $T=190$~mK. Dashed lines indicate full and $1/3$ pseudospin polarization. (b) False-color plot of specific heat $C/T$ measured as a function of temperature and field. The sharp high-temperature lambda anomaly (dotted line) discussed in \cite{Zhong2020} has an endpoint (diamond). The dashed line is a broad crossover. The solid line encloses a phase pocket with an ordering temperature $T_c$. At $T\rightarrow 0$ there is a quantum critical point at $H_c$. Circles indicate the positions of local specific heat maxima in temperature scans measured at a constant field.}\label{bulk}
\end{figure}

Our first finding is that the magnetic state above 1~T is {\em not} a fully polarized one, but an $m=1/3$ magnetization plateau. This is obvious from the magnetization curves measured at $T=1.3$~K on a 5.4~mg single-crystal sample by the coaxial pick-up coil pulsed-field magnetometer at HLD-EMFL \cite{Dresdenmagnetometer},
 and plotted in Fig.~\ref{bulk}a. As in all experiments reported below the field is applied along the crystallographic $c$-axis (magnetic easy axis). Note that full saturation is reached only at $\mu_0H_\text{sat}\approx 22$~T. The difference between the up-sweep and down-sweep curves is instrumental and due to magnetocaloric effect  \cite{supp}. The same plot shows Faraday balance magnetometry data \cite{Blosser2020} collected at $190$~mK. The low-temperature value for the start of the plateau phase is $\mu_0H_c\approx 0.8(1)$~T. Assuming the nearest-neighbor XXZ model, we can link $H_\text{sat}$ to the exchange constants for the in- and out-of-plane spin components: $g\mu_B\mu_0H_\text{sat}= 3S(J_{xy} + 2J_{zz})$, an exact SWT result. Previous theoretical work relates the ratio $\alpha\equiv J_{xy}/J_{zz}$ to $H_\text{sat}/H_c$ \cite{Miyashita1986,Yamamoto2014}. This allows us to estimate $J_{zz}= 3.1(3)$~meV and $\alpha=0.08(1)$ from the two measured transition field values.

 A more detailed phase diagram of \KSeCo was established in  calorimetric experiments  on a Quantum Design PPMS with a dilution refrigerator insert using a 0.05~mg single-crystal sample. The result is presented in a $C/T$ specific  heat plot in Fig.~\ref{bulk}b  \cite{supp}. The sharp high-temperature lambda-anomaly reported in \cite{Zhong2020} is also visible in our data (dotted line). This phase boundary has a well-defined endpoint at ($\mu_0 H_0\approx1.05$~T, $T_0\approx 4.5$~K). At lower field this anomaly is absent. Instead there is a broad crossover line (dashed) that extends down to zero field and corresponds to the bump at $T \sim 1$~K. An additional phase boundary marked by somewhat rounded specific heat maxima is detected at low temperatures. It  has the shape of a ``dome'' of some ordered phase with a zero-field ordering temperature $T_c\approx 0.35$~K, and terminates in a quantum critical point $\mu_0H_\text{c}=0.8$~T{, {\em i.e.}, right at the observed boundary of the magnetization plateau}. Overall, the phase diagram is qualitatively similar to that theoretically predicted for Na$_2$BaCo(PO$_4$)$_2$, where the sharp lambda-anomaly in field is a Potts transition to the uud phase,  replaced by broad  features due to BKT transitions at low fields \cite{Gao2022}.

\begin{figure}
	\includegraphics[width=\columnwidth]{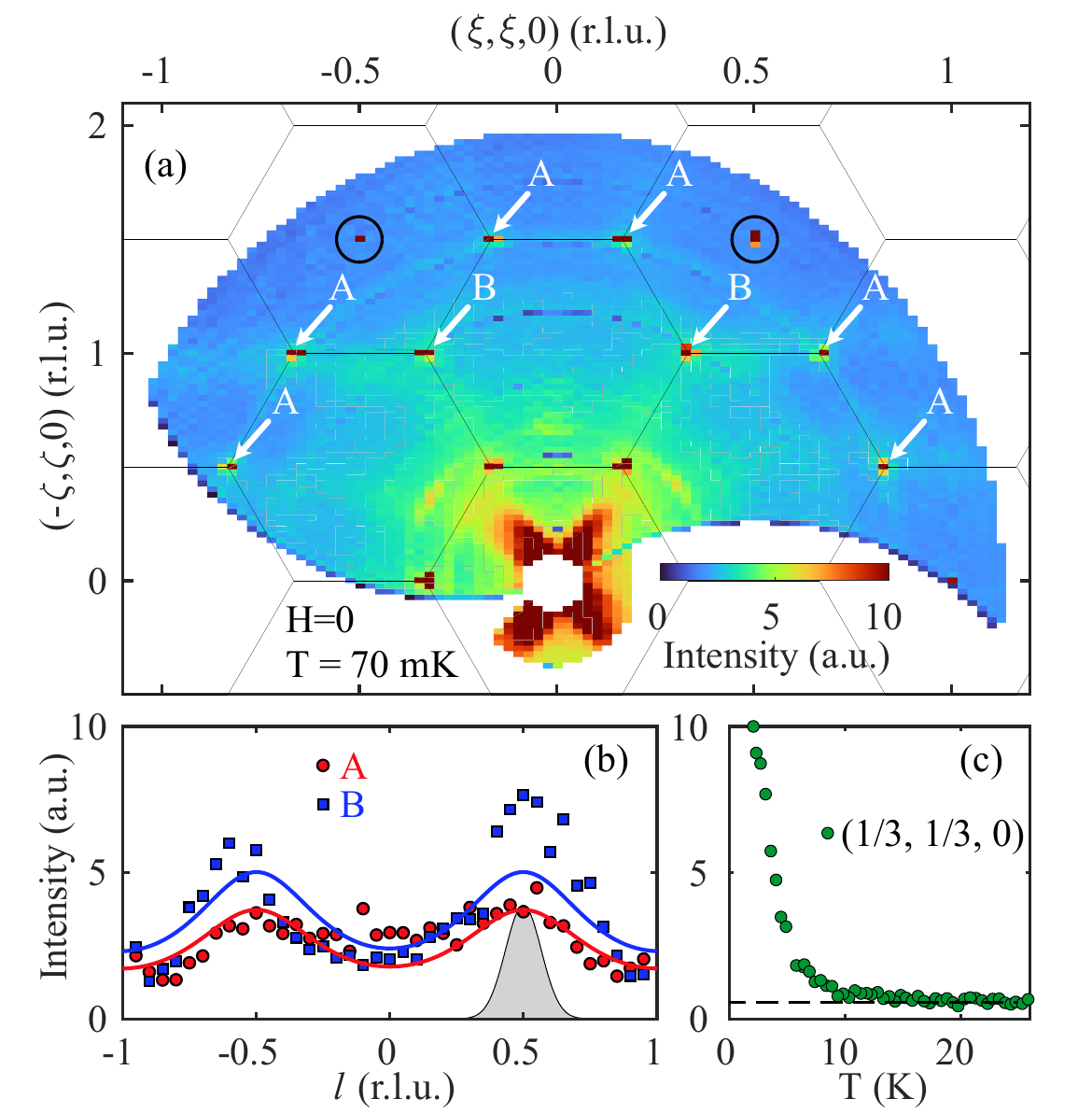}
	\caption{Elastic scattering measured in \KSeCo in zero field. (a) TOF data at $T=70$~mK in the $(h,k,0)$ plane reveal $(1/3,1/3)$-type two-dimensional order (arrows). Circles are structural Bragg peaks. Solid lines are 2D Brillouin zone boundaries for a single triangular plane. ``A'' and ``B'' are two sets of equivalent Bragg points that differ only by the absolute value of scattering vector. (b) Background-subtracterd $l$-scans averaged over each set. Solid lines are fits as described in the text. The shaded Gaussian represents experimental resolution as measured on the (-2,1,0) Bragg peak. (c) Temperature dependence of diffraction intensity at $(1/3,1/3,0)$ measured on ZEBRA. The dashed line is the background level measured just off the $(1/3,1/3)$ position.  }\label{diff}
\end{figure}

Next we performed neutron diffraction experiments at the DMC  powder instrument  and ZEBRA  lifting-counter diffractometer at Paul Scherrer Insitut, as well as the AMATERAS time-of-flight (TOF) spectrometer at J-PARC \cite{AMATERAS} . In the latter we utilized 75 and 425~mg single-crystal samples, respectively. Sample environment was a $^3$He-$^4$He dilution refrigerator on AMATERAS and a standard $^4$He cryostat on ZEBRA.  In both cases a cryomagnet was used to produce a magnetic field along the $c$ axis. Elastic scattering intensity measured at $T=70$~mK in zero field in the $(h,k,0)$ plane using the TOF technique with $E_\text{i}=7.73$~meV incident-energy neutrons is shown in Fig.~\ref{diff}a. A practically identical diffraction pattern is seen in a $\mu_0H=1.5$~T field \cite{supp}. Both demonstrate magnetic Bragg scattering indexed by  a $(1/3,1/3)$ 2D propagation vector (arrows). This corresponds to a $\sqrt{3} \times \sqrt{3}$ magnetic unit cell. In zero field this scattering sets in without any apparent phase transition (Fig.~\ref{diff}c, ZEBRA data), and simply increases steadily upon cooling below about 10~K. At higher fields it follows a more typical order-parameter-like curve with a transition exactly corresponding to the specific heat lambda-anomaly  \cite{supp}. 

{ The observed magnetic scattering is resolution-limited in the $(h,k)$ plane, indicating long-range order in both cases.  Fig.~\ref{diff}b show $l$-scans averaged over two sets of equivalent Bragg reflections  \cite{supp}.  The very broad peaks indicate only short-range correlations along $c$, of approximately AF character. The scans were fit to pairs of Voigt functions centered at $l=\pm 1/2$ (solid lines in Fig.~\ref{diff}b), the Gaussian component fixed at the experimental resolution \cite{supp}. From the Lorentzian width we extract the inter-plane correlation length $\lambda \approx 11(1)$ \AA, or just 0.6 lattice units. At $\mu_0 H =1.5$~T the correlation length is only slightly larger \cite{supp}. A more detailed analysis is hardly possible given the narrow $l$-coverage.}

{
Only two sets of equivalent Bragg positions are reachable outside of the strong  background region at low-$q$. The $uud$ and supersolid phases share the same propagation vector, a determination of the 2D magnetic structures does not appear feasible. To get a rough idea of the ordered moment in zero field, we compared the fitted Lorentzian intensities to those determined in $\mu_0H=1.5$~T and assumed the latter to be a $uud$ colinear structure. This procedure yields the estimate 2.6(3) $\mu_\text{B}$ per Co$^{2+}$, substantially reduced compared to 3.95 $\mu_\text{B}$ at saturation (Fig.~\ref{bulk}a).}

We can seek further guidance from theory. The ground state of the {\em classical} XXZ easy-axis triangular antiferromagnet is a 3-sublattice planar ``Y'' structure \cite{Miyashita1986}, as shown in Fig.~\ref{struc}b.  All spins on one sublattice are pointing ``down''. On the two other sublattices they are canted away from the ``up'' direction forming an angle $\phi<2\pi/3$ between them. This is a spin density wave for the $z$-components with a $\sqrt{3}\times \sqrt{3}$ unit cell, the ``down'' spins forming a honeycomb. It thereby breaks a $\mathbb{Z}_3$ discrete translation symmetry.  Crucially, the in-plane spin components of the Y-structure also breaks continuous $SO(2)$ rotational symmetry around the $z$ axis, which allows for gapless Goldstone modes. 

In applied fields theory predicts a transition to the colinear $uud$ plateau phase illustrated in Fig.~\ref{struc}c. Here rotational symmetry around the $z$ axis is preserved and the magnetic excitation spectrum is gapped. Upon decreasing the field, the transition to the Y-phase occurs when the lowest gapped magnon softens at some critical field $H_c$. As realized by  Batyev and Braginski sense \cite{Batyev1984} this transition is mapped on a Bose-Einstein condensation of magnons. Since it occurs in a ``spin-solid'' state with spontaneously broken translation symmetry, the low-field state is a ``spin-supersolid'' in the Andreev sense \cite{Andreev1971, Starykh2015}.

For quantum spins the situation is qualitatively similar \cite{Heidarian2010,Yamamoto2014}. This prompts us to hypothesize the following picture, that in many ways echoes the behavior predicted theoretically for  Na$_2$BaCo(PO$_4$)$_2$ \cite{Gao2022}. The robust $(1/3, 1/3)$ 2D magnetic correlations observed experimentally in \KSeCo are primarily due to the $z$ components of spins, which are similarly ordered in both $uud$- and Y-phases. They form already at $T\gg T_c$ and represent the spin solid. In zero field $T_c$ then corresponds to long-range or BKT ordering of the transverse spin components, resulting in a compressible spin-supersolid with a gapless spectrum. A BEC-type solid-supersolid quantum phase transition occurs at $H_c$.  The main differences with the theory on Na$_2$BaCo(PO$_4$)$_2$ are the presence of the end-point at $(H_0,T_0)$ in \KSeCo, a much wider plateau, and a much higher saturation field.

\begin{figure*}
	\includegraphics[width=\textwidth]{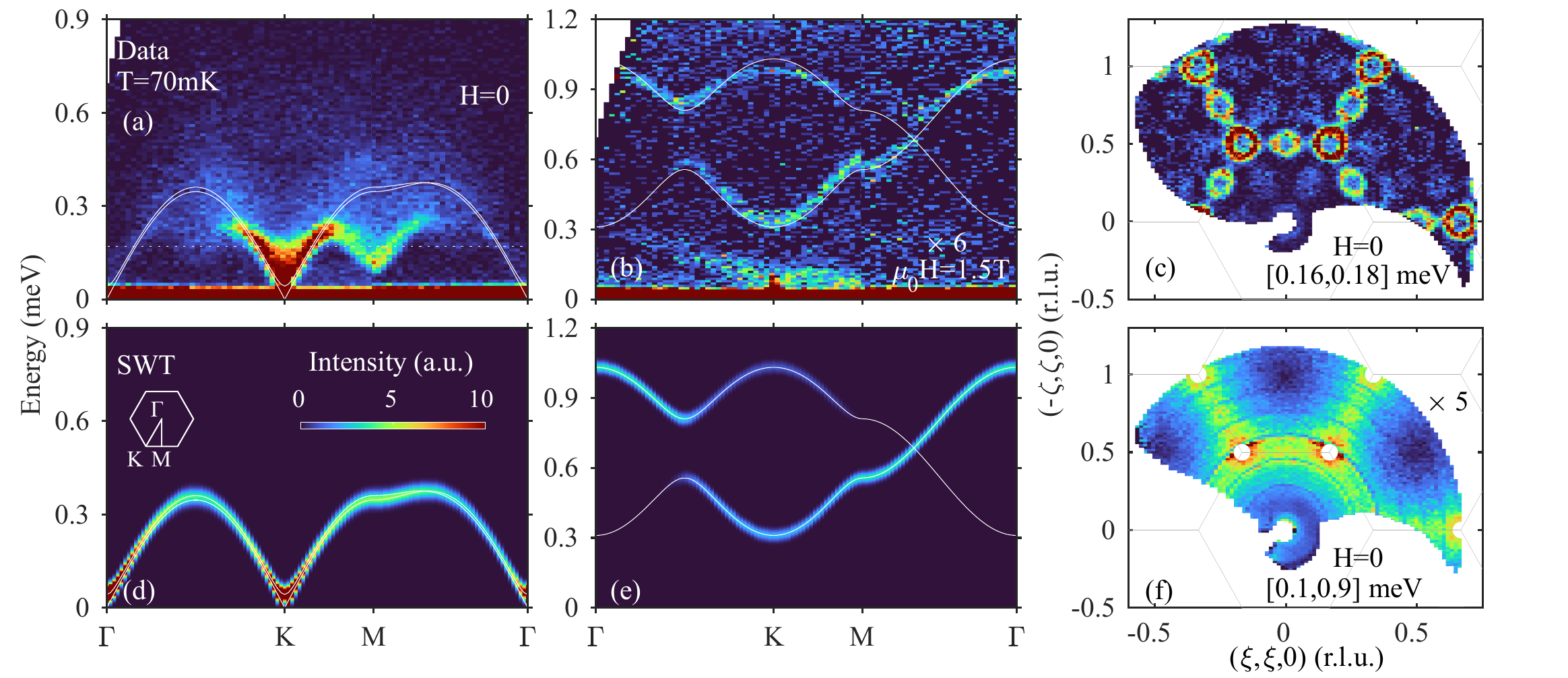}
	\caption{False-color plots of inelastic neutron intensity measured in \KSeCo at $T=70$~mK in zero field (a and c), and in $\mu_0H=1.5$~T (b, note the $\times 6$ scale factor). No background is subtracted. The low-energy scattering below 0.3 meV in (b) is spurious, originating from the sample environment. It is also present in (a) but not visible due to a different color range.  The dashed horizontal line in (a) is the position of the constant-energy cut in (c).
		Panels (d,e) as well as white lines in (a,b) show simulations based on linear SWT. (f) Equal-time structure factor obtained by integrating the intensity in the range 0.1-0.9~meV, chosen to avoid elastic line and any background at higher energies where no obvious signal originating from the lower spectral branches is present.}\label{spectra}
\end{figure*}

The emerging picture is consistent with measurements of the magnetic excitations. In Fig.~\ref{spectra} we show false-color neutron intensity plots of energy-momentum cuts (a,b) and energy-integrated cuts (c,f) through the data collected at the AMATERAS spectrometer using $E_i=2.24$~meV neutrons. More cuts are shown in the Supplement, where the integration procedure is also explained. The data in (a, c and f) are measured in zero field. Fig.~\ref{spectra}b shows that measured at $\mu_0H=1.5$~T.  The difference between zero field and 1.5~T data could hardly be more vivid. In the former case excitations are gapless. The spectrum is a broad continuum with no coherent modes, reminiscent of multi-spinon continua in quantum spin chains \cite{Stone2003}. In contrast, at 1.5~T the excitations are all  resolution limited.

To emphasize the contrast, we compare the observations with linear SWT calculations.  We based these on a simple nearest-neighbor XXZ model with the exchange parameters as quoted above. All simulations were performed using the SPIN-W software package \cite{TothLake_JPCM_2015_SpinW}.   For the zero-field case, we assumed a tiny $H=30$~G field along the $c$ axis to overcome the accidental classical degeneracy and stabilize the assumed $Y$-structure \cite{supp}.  For calculations in applied fields we estimated the gyromagnetic ratio $g=7.9$ from the saturation magnetization. The resulting  dispersion is shown as white lines in Fig.~\ref{spectra}a,b. The corresponding intensity simulations are shown in Fig.~\ref{spectra}d,e \cite{supp}, assuming  a Gaussian broadening to match experimental energy resolution.

For $\mu_0H=1.5$~T SWT reproduces the data up to 1~meV transfer rather well. A small discrepancy remains in the curvature of the higher-energy branch  (see Supplement for a more quantitative discussion). Evidently, quantum corrections to linear-SWT remain significant. Similar effects were observed in the plateau phases of triangular-lattice materials such as Ba$_3$CoSb$_2$O$_9$ \cite{Ma2016,Kamiya2018,Itoh2017}.

In contrast, in zero field the computed low-energy spectra are drastically different from what is actually observed, and not only in what concerns excitation widths. For spinon continua in Heisenberg spin chains \cite{Muller1981}, the computed shape of the SWT dispersion qualitatively resembles the lower bound of the continuum. { Quite to the opposite, in \KSeCo the latter has a distinct dip at the $M$-point, similar to the ``roton-like minima'' in the easy-plane case \cite{Coldea2020}.  In \KSeCo, near the $M$-point,  the lower bound of the continuum shows a  ``relativistic'' dispersion  with a mass gap $\Delta_M=0.12$~meV. This is particulary clear from the constant-energy cut shown Fig.~\ref{spectra}c.  Weak replicas of the scattering at the $K$ and $M$ points are seen throughout the zone, revealing a hidden $\sqrt{12}\times \sqrt{12}$ periodicity in the dynamic correlations. The large $M$-point intensity contributes to the equal-time structure factor (Fig.~\ref{spectra}f) being more or less evenly distributed along the zone boundary. Overall, our measured low-energy spectra are in excellent agreement with recent numerical simulations for \KSeCo \cite{Ulaga2024,Xu2024}.

One possible explanation for the observed behavior is spin wave instability \cite{ZhitomirskyChernyshev_RMP_2013_DecayReview}, which is allowed in the non-colinear ``Y''- structure. This paradigm applies well to systems with rather isotropic interactions, such as Ba$_3$MnSb$_2$O$_9$ \cite{Mingfang2023} or Na$_2$BaCo(PO$_4$)$_2$ \cite{sheng2024}. In \KSeCo it is able to account for at least some observed features. In zero field the lower bound of the two-magnon continuum in SWT coincides with the lowest-energy magnon mode.  The higher-energy spin wave branch is thus always unstable towards two-particle decay and my give rise to a continuum. Unfortunately, this  provides no explanation for the energy minimum seen at the $M$-point.  Moreover, the SWT language may be totally inadequate for small $\alpha$.  In this limit theory predicts only a tiny transverse order parameter \cite{Wessel2005}. Acoustic spin waves are the corresponding Goldstone modes, and are nothing other than its precessions. Therefore, they can't be expected to contribute significantly to the spectrum.

A completely different view links the continuum in the spin-supersolid phase to excitations in a proximate Dirac spin liquid \cite{jia2023}. In this interpretation the continuum is due to multi-spinon, rather than multi-magnon states. The predicted continuum is bounded by a ``relativistic'' linear-dispersion cone not only near the $K$-point, but also near the $M$ and $\Gamma$ points, exactly as observed in \KSeCo. The theoretical intensity pattern for a constant-energy cut looks remarkably similar to what we see experimentally (Fig.~\ref{spectra}c). We can speculate that in the high-anisotropy limit  the spinon picture may provide a more adequate description of the physics.

In summary, the Ising-like triangular-lattice antiferromagnet \KSeCo, despite long-range order, demonstrates gapless continuum spin dynamics in the spin-supersolid phase. It strongly resembles that of deconfined spinons and may be indicative of a proximate Dirac spin liquid state.  }

\begin{acknowledgements} This work was Supported by a MINT grant of the Swiss National Science Foundation. AZ thanks Prof. A. Chernyshev, Dr. M. Zhitomirsky and Prof. P. Prelovsek for providing helpful feedback. Many thanks to J. Nagl for help with graphical illustrations. We acknowledge the support of the HLD at HZDR, member of the European Magnetic Field Laboratory (EMFL) , and the W\"urzburg-Dresden Cluster of Excellence on Complexity and Topology in Quantum Matter - ct.qmat (EXC 2147, project ID 390858490). Data at J-PARC were collected in Experiment no. 2023B0161.  While preparing this manuscript, we became aware of Ref.~\cite{Collin} that seems to announce independent but similar findings.

\end{acknowledgements}

\bibliography{KSeCo_19_08_2024}

\begin{thebibliography}{36}%
\makeatletter
\providecommand \@ifxundefined [1]{%
 \@ifx{#1\undefined}
}%
\providecommand \@ifnum [1]{%
 \ifnum #1\expandafter \@firstoftwo
 \else \expandafter \@secondoftwo
 \fi
}%
\providecommand \@ifx [1]{%
 \ifx #1\expandafter \@firstoftwo
 \else \expandafter \@secondoftwo
 \fi
}%
\providecommand \natexlab [1]{#1}%
\providecommand \enquote  [1]{``#1''}%
\providecommand \bibnamefont  [1]{#1}%
\providecommand \bibfnamefont [1]{#1}%
\providecommand \citenamefont [1]{#1}%
\providecommand \href@noop [0]{\@secondoftwo}%
\providecommand \href [0]{\begingroup \@sanitize@url \@href}%
\providecommand \@href[1]{\@@startlink{#1}\@@href}%
\providecommand \@@href[1]{\endgroup#1\@@endlink}%
\providecommand \@sanitize@url [0]{\catcode `\\12\catcode `\$12\catcode `\&12\catcode `\#12\catcode `\^12\catcode `\_12\catcode `\%12\relax}%
\providecommand \@@startlink[1]{}%
\providecommand \@@endlink[0]{}%
\providecommand \url  [0]{\begingroup\@sanitize@url \@url }%
\providecommand \@url [1]{\endgroup\@href {#1}{\urlprefix }}%
\providecommand \urlprefix  [0]{URL }%
\providecommand \Eprint [0]{\href }%
\providecommand \doibase [0]{https://doi.org/}%
\providecommand \selectlanguage [0]{\@gobble}%
\providecommand \bibinfo  [0]{\@secondoftwo}%
\providecommand \bibfield  [0]{\@secondoftwo}%
\providecommand \translation [1]{[#1]}%
\providecommand \BibitemOpen [0]{}%
\providecommand \bibitemStop [0]{}%
\providecommand \bibitemNoStop [0]{.\EOS\space}%
\providecommand \EOS [0]{\spacefactor3000\relax}%
\providecommand \BibitemShut  [1]{\csname bibitem#1\endcsname}%
\let\auto@bib@innerbib\@empty
\bibitem [{\citenamefont {Wannier}(1950)}]{Wannier1950}%
  \BibitemOpen
  \bibfield  {author} {\bibinfo {author} {\bibfnamefont {G.~H.}\ \bibnamefont {Wannier}},\ }\bibfield  {title} {\bibinfo {title} {Antiferromagnetism. the triangular {Ising} net},\ }\href {https://doi.org/10.1103/PhysRev.79.357} {\bibfield  {journal} {\bibinfo  {journal} {Phys. Rev.}\ }\textbf {\bibinfo {volume} {79}},\ \bibinfo {pages} {357} (\bibinfo {year} {1950})}\BibitemShut {NoStop}%
\bibitem [{\citenamefont {Miyashita}(1986)}]{Miyashita1986}%
  \BibitemOpen
  \bibfield  {author} {\bibinfo {author} {\bibfnamefont {S.}~\bibnamefont {Miyashita}},\ }\bibfield  {title} {\bibinfo {title} {Magnetic properties of {Ising}-like {Heisenberg} antiferromagnets on the triangular lattice},\ }\href {https://doi.org/10.1143/JPSJ.55.3605} {\bibfield  {journal} {\bibinfo  {journal} {Journal of the Physical Society of Japan}\ }\textbf {\bibinfo {volume} {55}},\ \bibinfo {pages} {3605} (\bibinfo {year} {1986})}\BibitemShut {NoStop}%
\bibitem [{\citenamefont {Heidarian}\ and\ \citenamefont {Paramekanti}(2010)}]{Heidarian2010}%
  \BibitemOpen
  \bibfield  {author} {\bibinfo {author} {\bibfnamefont {D.}~\bibnamefont {Heidarian}}\ and\ \bibinfo {author} {\bibfnamefont {A.}~\bibnamefont {Paramekanti}},\ }\bibfield  {title} {\bibinfo {title} {Supersolidity in the triangular lattice spin-$1/2$ {XXZ} model: A variational perspective},\ }\href {https://doi.org/10.1103/PhysRevLett.104.015301} {\bibfield  {journal} {\bibinfo  {journal} {Phys. Rev. Lett.}\ }\textbf {\bibinfo {volume} {104}},\ \bibinfo {pages} {015301} (\bibinfo {year} {2010})}\BibitemShut {NoStop}%
\bibitem [{\citenamefont {Yamamoto}\ \emph {et~al.}(2014)\citenamefont {Yamamoto}, \citenamefont {Marmorini},\ and\ \citenamefont {Danshita}}]{Yamamoto2014}%
  \BibitemOpen
  \bibfield  {author} {\bibinfo {author} {\bibfnamefont {D.}~\bibnamefont {Yamamoto}}, \bibinfo {author} {\bibfnamefont {G.}~\bibnamefont {Marmorini}},\ and\ \bibinfo {author} {\bibfnamefont {I.}~\bibnamefont {Danshita}},\ }\bibfield  {title} {\bibinfo {title} {Quantum phase diagram of the triangular-lattice {XXZ} model in a magnetic field},\ }\href {https://doi.org/10.1103/PhysRevLett.112.127203} {\bibfield  {journal} {\bibinfo  {journal} {Phys. Rev. Lett.}\ }\textbf {\bibinfo {volume} {112}},\ \bibinfo {pages} {127203} (\bibinfo {year} {2014})}\BibitemShut {NoStop}%
\bibitem [{\citenamefont {Gao}\ \emph {et~al.}(2022)\citenamefont {Gao}, \citenamefont {Fan}, \citenamefont {Li}, \citenamefont {Yang}, \citenamefont {Zeng}, \citenamefont {Sheng}, \citenamefont {Zhong}, \citenamefont {Qi}, \citenamefont {Wan},\ and\ \citenamefont {Li}}]{Gao2022}%
  \BibitemOpen
  \bibfield  {author} {\bibinfo {author} {\bibfnamefont {Y.}~\bibnamefont {Gao}}, \bibinfo {author} {\bibfnamefont {Y.-C.}\ \bibnamefont {Fan}}, \bibinfo {author} {\bibfnamefont {H.}~\bibnamefont {Li}}, \bibinfo {author} {\bibfnamefont {F.}~\bibnamefont {Yang}}, \bibinfo {author} {\bibfnamefont {X.-T.}\ \bibnamefont {Zeng}}, \bibinfo {author} {\bibfnamefont {X.-L.}\ \bibnamefont {Sheng}}, \bibinfo {author} {\bibfnamefont {R.}~\bibnamefont {Zhong}}, \bibinfo {author} {\bibfnamefont {Y.}~\bibnamefont {Qi}}, \bibinfo {author} {\bibfnamefont {Y.}~\bibnamefont {Wan}},\ and\ \bibinfo {author} {\bibfnamefont {W.}~\bibnamefont {Li}},\ }\bibfield  {title} {\bibinfo {title} {Spin supersolidity in nearly ideal easy-axis triangular quantum antiferromagnet {Na$_2$BaCo(PO$_4$)$_2$}},\ }\href {https://doi.org/10.1038/s41535-022-00500-3} {\bibfield  {journal} {\bibinfo  {journal} {npj Quantum Materials}\ }\textbf {\bibinfo {volume} {7}},\ \bibinfo {pages} {89} (\bibinfo {year} {2022})}\BibitemShut {NoStop}%
\bibitem [{\citenamefont {Ma}\ \emph {et~al.}(2016)\citenamefont {Ma}, \citenamefont {Kamiya}, \citenamefont {Hong}, \citenamefont {Cao}, \citenamefont {Ehlers}, \citenamefont {Tian}, \citenamefont {Batista}, \citenamefont {Dun}, \citenamefont {Zhou},\ and\ \citenamefont {Matsuda}}]{Ma2016}%
  \BibitemOpen
  \bibfield  {author} {\bibinfo {author} {\bibfnamefont {J.}~\bibnamefont {Ma}}, \bibinfo {author} {\bibfnamefont {Y.}~\bibnamefont {Kamiya}}, \bibinfo {author} {\bibfnamefont {T.}~\bibnamefont {Hong}}, \bibinfo {author} {\bibfnamefont {H.~B.}\ \bibnamefont {Cao}}, \bibinfo {author} {\bibfnamefont {G.}~\bibnamefont {Ehlers}}, \bibinfo {author} {\bibfnamefont {W.}~\bibnamefont {Tian}}, \bibinfo {author} {\bibfnamefont {C.~D.}\ \bibnamefont {Batista}}, \bibinfo {author} {\bibfnamefont {Z.~L.}\ \bibnamefont {Dun}}, \bibinfo {author} {\bibfnamefont {H.~D.}\ \bibnamefont {Zhou}},\ and\ \bibinfo {author} {\bibfnamefont {M.}~\bibnamefont {Matsuda}},\ }\bibfield  {title} {\bibinfo {title} {Static and dynamical properties of the spin-$1/2$ equilateral triangular-lattice antiferromagnet {${\mathrm{Ba}}_{3}{\mathrm{CoSb}}_{2}{\mathrm{O}}_{9}$}},\ }\href {https://doi.org/10.1103/PhysRevLett.116.087201} {\bibfield  {journal} {\bibinfo  {journal} {Phys. Rev. Lett.}\ }\textbf {\bibinfo {volume} {116}},\ \bibinfo {pages}
  {087201} (\bibinfo {year} {2016})}\BibitemShut {NoStop}%
\bibitem [{\citenamefont {Kamiya}\ \emph {et~al.}(2018)\citenamefont {Kamiya}, \citenamefont {Ge}, \citenamefont {Hong}, \citenamefont {Qiu}, \citenamefont {Quintero-Castro}, \citenamefont {Lu}, \citenamefont {Cao}, \citenamefont {Matsuda}, \citenamefont {Choi}, \citenamefont {Batista}, \citenamefont {Mourigal}, \citenamefont {D.},\ and\ \citenamefont {Ma}}]{Kamiya2018}%
  \BibitemOpen
  \bibfield  {author} {\bibinfo {author} {\bibfnamefont {Y.}~\bibnamefont {Kamiya}}, \bibinfo {author} {\bibfnamefont {L.}~\bibnamefont {Ge}}, \bibinfo {author} {\bibfnamefont {T.}~\bibnamefont {Hong}}, \bibinfo {author} {\bibfnamefont {Y.}~\bibnamefont {Qiu}}, \bibinfo {author} {\bibfnamefont {D.~L.}\ \bibnamefont {Quintero-Castro}}, \bibinfo {author} {\bibfnamefont {Z.}~\bibnamefont {Lu}}, \bibinfo {author} {\bibfnamefont {H.~B.}\ \bibnamefont {Cao}}, \bibinfo {author} {\bibfnamefont {M.}~\bibnamefont {Matsuda}}, \bibinfo {author} {\bibfnamefont {E.~S.}\ \bibnamefont {Choi}}, \bibinfo {author} {\bibfnamefont {C.~D.}\ \bibnamefont {Batista}}, \bibinfo {author} {\bibfnamefont {M.}~\bibnamefont {Mourigal}}, \bibinfo {author} {\bibfnamefont {Z.~H.}\ \bibnamefont {D.}},\ and\ \bibinfo {author} {\bibfnamefont {J.}~\bibnamefont {Ma}},\ }\bibfield  {title} {\bibinfo {title} {{The nature of spin excitations in the one-third magnetization plateau phase of Ba$_3$CoSb$_2$O$_9$}},\ }\href
  {https://doi.org/10.1038/s41467-018-04914-1} {\bibfield  {journal} {\bibinfo  {journal} {Nat. Commun.}\ }\textbf {\bibinfo {volume} {9}},\ \bibinfo {pages} {2666} (\bibinfo {year} {2018})}\BibitemShut {NoStop}%
\bibitem [{\citenamefont {Ito}\ \emph {et~al.}(2017)\citenamefont {Ito}, \citenamefont {Kurita}, \citenamefont {Tanaka}, \citenamefont {Ohira-Kawamura}, \citenamefont {Nakajima}, \citenamefont {Itoh}, \citenamefont {Kuwahara},\ and\ \citenamefont {Kakurai}}]{Itoh2017}%
  \BibitemOpen
  \bibfield  {author} {\bibinfo {author} {\bibfnamefont {S.}~\bibnamefont {Ito}}, \bibinfo {author} {\bibfnamefont {N.}~\bibnamefont {Kurita}}, \bibinfo {author} {\bibfnamefont {H.}~\bibnamefont {Tanaka}}, \bibinfo {author} {\bibfnamefont {S.}~\bibnamefont {Ohira-Kawamura}}, \bibinfo {author} {\bibfnamefont {K.}~\bibnamefont {Nakajima}}, \bibinfo {author} {\bibfnamefont {S.}~\bibnamefont {Itoh}}, \bibinfo {author} {\bibfnamefont {K.}~\bibnamefont {Kuwahara}},\ and\ \bibinfo {author} {\bibfnamefont {K.}~\bibnamefont {Kakurai}},\ }\bibfield  {title} {\bibinfo {title} {{Structure of the magnetic excitations in the spin-$1/2$ triangular-lattice Heisenberg antiferromagnet Ba$_3$CoSb$_2$O$_9$}},\ }\href {https://doi.org/10.1038/s41467-017-00316-x} {\bibfield  {journal} {\bibinfo  {journal} {Nat. Comm.}\ }\textbf {\bibinfo {volume} {8}},\ \bibinfo {pages} {235} (\bibinfo {year} {2017})}\BibitemShut {NoStop}%
\bibitem [{\citenamefont {Macdougal}\ \emph {et~al.}(2020)\citenamefont {Macdougal}, \citenamefont {Williams}, \citenamefont {Prabhakaran}, \citenamefont {Bewley}, \citenamefont {Voneshen},\ and\ \citenamefont {Coldea}}]{Coldea2020}%
  \BibitemOpen
  \bibfield  {author} {\bibinfo {author} {\bibfnamefont {D.}~\bibnamefont {Macdougal}}, \bibinfo {author} {\bibfnamefont {S.}~\bibnamefont {Williams}}, \bibinfo {author} {\bibfnamefont {D.}~\bibnamefont {Prabhakaran}}, \bibinfo {author} {\bibfnamefont {R.~I.}\ \bibnamefont {Bewley}}, \bibinfo {author} {\bibfnamefont {D.~J.}\ \bibnamefont {Voneshen}},\ and\ \bibinfo {author} {\bibfnamefont {R.}~\bibnamefont {Coldea}},\ }\bibfield  {title} {\bibinfo {title} {Avoided quasiparticle decay and enhanced excitation continuum in the spin-$\frac{1}{2}$ near-{Heisenberg} triangular antiferromagnet {${\mathrm{Ba}}_{3}{\mathrm{CoSb}}_{2}{\mathrm{O}}_{9}$}},\ }\href {https://doi.org/10.1103/PhysRevB.102.064421} {\bibfield  {journal} {\bibinfo  {journal} {Phys. Rev. B}\ }\textbf {\bibinfo {volume} {102}},\ \bibinfo {pages} {064421} (\bibinfo {year} {2020})}\BibitemShut {NoStop}%
\bibitem [{\citenamefont {Ghioldi}\ \emph {et~al.}(2022)\citenamefont {Ghioldi}, \citenamefont {Zhang}, \citenamefont {Kamiya}, \citenamefont {Manuel}, \citenamefont {Trumper},\ and\ \citenamefont {Batista}}]{Ghioldi2022}%
  \BibitemOpen
  \bibfield  {author} {\bibinfo {author} {\bibfnamefont {E.~A.}\ \bibnamefont {Ghioldi}}, \bibinfo {author} {\bibfnamefont {S.-S.}\ \bibnamefont {Zhang}}, \bibinfo {author} {\bibfnamefont {Y.}~\bibnamefont {Kamiya}}, \bibinfo {author} {\bibfnamefont {L.~O.}\ \bibnamefont {Manuel}}, \bibinfo {author} {\bibfnamefont {A.~E.}\ \bibnamefont {Trumper}},\ and\ \bibinfo {author} {\bibfnamefont {C.~D.}\ \bibnamefont {Batista}},\ }\bibfield  {title} {\bibinfo {title} {{Evidence of two-spinon bound states in the magnetic spectrum of ${\mathrm{Ba}}_{3}{\mathrm{CoSb}}_{2}{\mathrm{O}}_{9}$}},\ }\href {https://doi.org/10.1103/PhysRevB.106.064418} {\bibfield  {journal} {\bibinfo  {journal} {Phys. Rev. B}\ }\textbf {\bibinfo {volume} {106}},\ \bibinfo {pages} {064418} (\bibinfo {year} {2022})}\BibitemShut {NoStop}%
\bibitem [{\citenamefont {Anderson}(1973)}]{Anderson1973}%
  \BibitemOpen
  \bibfield  {author} {\bibinfo {author} {\bibfnamefont {P.}~\bibnamefont {Anderson}},\ }\bibfield  {title} {\bibinfo {title} {Resonating valence bonds: A new kind of insulator?},\ }\href {https://doi.org/http://dx.doi.org/10.1016/0025-5408(73)90167-0} {\bibfield  {journal} {\bibinfo  {journal} {Materials Research Bulletin}\ }\textbf {\bibinfo {volume} {8}},\ \bibinfo {pages} {153 } (\bibinfo {year} {1973})}\BibitemShut {NoStop}%
\bibitem [{\citenamefont {Balents}(2010)}]{Balents2010}%
  \BibitemOpen
  \bibfield  {author} {\bibinfo {author} {\bibfnamefont {L.}~\bibnamefont {Balents}},\ }\bibfield  {title} {\bibinfo {title} {Spin liquids in frustrated magnets},\ }\href {https://doi.org/10.1038/nature08917} {\bibfield  {journal} {\bibinfo  {journal} {Nature}\ }\textbf {\bibinfo {volume} {464}},\ \bibinfo {pages} {199 EP } (\bibinfo {year} {2010})}\BibitemShut {NoStop}%
\bibitem [{\citenamefont {Iqbal}\ \emph {et~al.}(2016)\citenamefont {Iqbal}, \citenamefont {Hu}, \citenamefont {Thomale}, \citenamefont {Poilblanc},\ and\ \citenamefont {Becca}}]{Iqbal2016}%
  \BibitemOpen
  \bibfield  {author} {\bibinfo {author} {\bibfnamefont {Y.}~\bibnamefont {Iqbal}}, \bibinfo {author} {\bibfnamefont {W.-J.}\ \bibnamefont {Hu}}, \bibinfo {author} {\bibfnamefont {R.}~\bibnamefont {Thomale}}, \bibinfo {author} {\bibfnamefont {D.}~\bibnamefont {Poilblanc}},\ and\ \bibinfo {author} {\bibfnamefont {F.}~\bibnamefont {Becca}},\ }\bibfield  {title} {\bibinfo {title} {Spin liquid nature in the {Heisenberg} ${J}_{1}\ensuremath{-}{J}_{2}$ triangular antiferromagnet},\ }\href {https://doi.org/10.1103/PhysRevB.93.144411} {\bibfield  {journal} {\bibinfo  {journal} {Phys. Rev. B}\ }\textbf {\bibinfo {volume} {93}},\ \bibinfo {pages} {144411} (\bibinfo {year} {2016})}\BibitemShut {NoStop}%
\bibitem [{\citenamefont {Wietek}\ and\ \citenamefont {L\"auchli}(2017)}]{Wietek2017}%
  \BibitemOpen
  \bibfield  {author} {\bibinfo {author} {\bibfnamefont {A.}~\bibnamefont {Wietek}}\ and\ \bibinfo {author} {\bibfnamefont {A.~M.}\ \bibnamefont {L\"auchli}},\ }\bibfield  {title} {\bibinfo {title} {Chiral spin liquid and quantum criticality in extended ${S}=\frac{1}{2}$ {Heisenberg} models on the triangular lattice},\ }\href {https://doi.org/10.1103/PhysRevB.95.035141} {\bibfield  {journal} {\bibinfo  {journal} {Phys. Rev. B}\ }\textbf {\bibinfo {volume} {95}},\ \bibinfo {pages} {035141} (\bibinfo {year} {2017})}\BibitemShut {NoStop}%
\bibitem [{\citenamefont {Hu}\ \emph {et~al.}(2019)\citenamefont {Hu}, \citenamefont {Zhu}, \citenamefont {Eggert},\ and\ \citenamefont {He}}]{Hu2019}%
  \BibitemOpen
  \bibfield  {author} {\bibinfo {author} {\bibfnamefont {S.}~\bibnamefont {Hu}}, \bibinfo {author} {\bibfnamefont {W.}~\bibnamefont {Zhu}}, \bibinfo {author} {\bibfnamefont {S.}~\bibnamefont {Eggert}},\ and\ \bibinfo {author} {\bibfnamefont {Y.-C.}\ \bibnamefont {He}},\ }\bibfield  {title} {\bibinfo {title} {Dirac spin liquid on the spin-$1/2$ triangular {Heisenberg} antiferromagnet},\ }\href {https://doi.org/10.1103/PhysRevLett.123.207203} {\bibfield  {journal} {\bibinfo  {journal} {Phys. Rev. Lett.}\ }\textbf {\bibinfo {volume} {123}},\ \bibinfo {pages} {207203} (\bibinfo {year} {2019})}\BibitemShut {NoStop}%
\bibitem [{\citenamefont {Zhong}\ \emph {et~al.}(2020)\citenamefont {Zhong}, \citenamefont {Guo},\ and\ \citenamefont {Cava}}]{Zhong2020}%
  \BibitemOpen
  \bibfield  {author} {\bibinfo {author} {\bibfnamefont {R.}~\bibnamefont {Zhong}}, \bibinfo {author} {\bibfnamefont {S.}~\bibnamefont {Guo}},\ and\ \bibinfo {author} {\bibfnamefont {R.~J.}\ \bibnamefont {Cava}},\ }\bibfield  {title} {\bibinfo {title} {Frustrated magnetism in the layered triangular lattice materials {${\mathrm{K}}_{2}\mathrm{Co}{({\mathrm{SeO}}_{3})}_{2}$ and ${\mathrm{Rb}}_{2}\mathrm{Co}{({\mathrm{SeO}}_{3})}_{2}$}},\ }\href {https://doi.org/10.1103/PhysRevMaterials.4.084406} {\bibfield  {journal} {\bibinfo  {journal} {Phys. Rev. Mater.}\ }\textbf {\bibinfo {volume} {4}},\ \bibinfo {pages} {084406} (\bibinfo {year} {2020})}\BibitemShut {NoStop}%
\bibitem [{\citenamefont {Wildner}(1992)}]{Wildner1994}%
  \BibitemOpen
  \bibfield  {author} {\bibinfo {author} {\bibfnamefont {M.}~\bibnamefont {Wildner}},\ }\bibfield  {title} {\bibinfo {title} {{Isotypism of a selenite with a carbonate: structure of the buetschliite-type compound K${\sb 2}$Co(SeO${\sb 3}$)${\sb 2}$}},\ }\href {https://doi.org/10.1107/S0108270191011459} {\bibfield  {journal} {\bibinfo  {journal} {Acta Crystallographica Section C}\ }\textbf {\bibinfo {volume} {48}},\ \bibinfo {pages} {410} (\bibinfo {year} {1992})}\BibitemShut {NoStop}%
\bibitem [{\citenamefont {Wellm}\ \emph {et~al.}(2021)\citenamefont {Wellm}, \citenamefont {Roscher}, \citenamefont {Zeisner}, \citenamefont {Alfonsov}, \citenamefont {Zhong}, \citenamefont {Cava}, \citenamefont {Savoyant}, \citenamefont {Hayn}, \citenamefont {van~den Brink}, \citenamefont {B\"uchner}, \citenamefont {Janson},\ and\ \citenamefont {Kataev}}]{Wellm2021}%
  \BibitemOpen
  \bibfield  {author} {\bibinfo {author} {\bibfnamefont {C.}~\bibnamefont {Wellm}}, \bibinfo {author} {\bibfnamefont {W.}~\bibnamefont {Roscher}}, \bibinfo {author} {\bibfnamefont {J.}~\bibnamefont {Zeisner}}, \bibinfo {author} {\bibfnamefont {A.}~\bibnamefont {Alfonsov}}, \bibinfo {author} {\bibfnamefont {R.}~\bibnamefont {Zhong}}, \bibinfo {author} {\bibfnamefont {R.~J.}\ \bibnamefont {Cava}}, \bibinfo {author} {\bibfnamefont {A.}~\bibnamefont {Savoyant}}, \bibinfo {author} {\bibfnamefont {R.}~\bibnamefont {Hayn}}, \bibinfo {author} {\bibfnamefont {J.}~\bibnamefont {van~den Brink}}, \bibinfo {author} {\bibfnamefont {B.}~\bibnamefont {B\"uchner}}, \bibinfo {author} {\bibfnamefont {O.}~\bibnamefont {Janson}},\ and\ \bibinfo {author} {\bibfnamefont {V.}~\bibnamefont {Kataev}},\ }\bibfield  {title} {\bibinfo {title} {Frustration enhanced by {K}itaev exchange in a $\tilde{j}_{\text{eff}}=\frac{1}{2}$ triangular antiferromagnet},\ }\href {https://doi.org/10.1103/PhysRevB.104.L100420} {\bibfield  {journal}
  {\bibinfo  {journal} {Phys. Rev. B}\ }\textbf {\bibinfo {volume} {104}},\ \bibinfo {pages} {L100420} (\bibinfo {year} {2021})}\BibitemShut {NoStop}%
\bibitem [{\citenamefont {Skourski}\ \emph {et~al.}(2011)\citenamefont {Skourski}, \citenamefont {Kuz'min}, \citenamefont {Skokov}, \citenamefont {Andreev},\ and\ \citenamefont {Wosnitza}}]{Dresdenmagnetometer}%
  \BibitemOpen
  \bibfield  {author} {\bibinfo {author} {\bibfnamefont {Y.}~\bibnamefont {Skourski}}, \bibinfo {author} {\bibfnamefont {M.~D.}\ \bibnamefont {Kuz'min}}, \bibinfo {author} {\bibfnamefont {K.~P.}\ \bibnamefont {Skokov}}, \bibinfo {author} {\bibfnamefont {A.~V.}\ \bibnamefont {Andreev}},\ and\ \bibinfo {author} {\bibfnamefont {J.}~\bibnamefont {Wosnitza}},\ }\bibfield  {title} {\bibinfo {title} {High-field magnetization of {Ho${}_{2}$Fe${}_{17}$}},\ }\href {https://doi.org/10.1103/PhysRevB.83.214420} {\bibfield  {journal} {\bibinfo  {journal} {Phys. Rev. B}\ }\textbf {\bibinfo {volume} {83}},\ \bibinfo {pages} {214420} (\bibinfo {year} {2011})}\BibitemShut {NoStop}%
\bibitem [{sup()}]{supp}%
  \BibitemOpen
  \href@noop {} {}\bibinfo {note} {{See supplementary Material.}}\BibitemShut {Stop}%
\bibitem [{\citenamefont {Blosser}\ \emph {et~al.}(2020)\citenamefont {Blosser}, \citenamefont {Facheris},\ and\ \citenamefont {Zheludev}}]{Blosser2020}%
  \BibitemOpen
  \bibfield  {author} {\bibinfo {author} {\bibfnamefont {D.}~\bibnamefont {Blosser}}, \bibinfo {author} {\bibfnamefont {L.}~\bibnamefont {Facheris}},\ and\ \bibinfo {author} {\bibfnamefont {A.}~\bibnamefont {Zheludev}},\ }\bibfield  {title} {\bibinfo {title} {{Miniature capacitive Faraday force magnetometer for magnetization measurements at low temperatures and high magnetic fields}},\ }\href {https://doi.org/10.1063/5.0005850} {\bibfield  {journal} {\bibinfo  {journal} {Review of Scientific Instruments}\ }\textbf {\bibinfo {volume} {91}},\ \bibinfo {pages} {073905} (\bibinfo {year} {2020})}\BibitemShut {NoStop}%
\bibitem [{\citenamefont {Nakajima}\ \emph {et~al.}(2011)\citenamefont {Nakajima}, \citenamefont {Ohira-Kawamura}, \citenamefont {Kikuchi}, \citenamefont {Nakamura}, \citenamefont {Kajimoto}, \citenamefont {Inamura}, \citenamefont {Takahashi}, \citenamefont {Aizawa}, \citenamefont {Suzuya}, \citenamefont {Shibata}, \citenamefont {Nakatani}, \citenamefont {Soyama}, \citenamefont {Maruyama}, \citenamefont {Tanaka}, \citenamefont {Kambara}, \citenamefont {Iwahashi}, \citenamefont {Itoh}, \citenamefont {Osakabe}, \citenamefont {Wakimoto}, \citenamefont {Kakurai}, \citenamefont {Maekawa}, \citenamefont {Harada}, \citenamefont {Oikawa}, \citenamefont {E.~Lechner}, \citenamefont {Mezei},\ and\ \citenamefont {Arai}}]{AMATERAS}%
  \BibitemOpen
  \bibfield  {author} {\bibinfo {author} {\bibfnamefont {K.}~\bibnamefont {Nakajima}}, \bibinfo {author} {\bibfnamefont {S.}~\bibnamefont {Ohira-Kawamura}}, \bibinfo {author} {\bibfnamefont {T.}~\bibnamefont {Kikuchi}}, \bibinfo {author} {\bibfnamefont {M.}~\bibnamefont {Nakamura}}, \bibinfo {author} {\bibfnamefont {R.}~\bibnamefont {Kajimoto}}, \bibinfo {author} {\bibfnamefont {Y.}~\bibnamefont {Inamura}}, \bibinfo {author} {\bibfnamefont {N.}~\bibnamefont {Takahashi}}, \bibinfo {author} {\bibfnamefont {K.}~\bibnamefont {Aizawa}}, \bibinfo {author} {\bibfnamefont {K.}~\bibnamefont {Suzuya}}, \bibinfo {author} {\bibfnamefont {K.}~\bibnamefont {Shibata}}, \bibinfo {author} {\bibfnamefont {T.}~\bibnamefont {Nakatani}}, \bibinfo {author} {\bibfnamefont {K.}~\bibnamefont {Soyama}}, \bibinfo {author} {\bibfnamefont {R.}~\bibnamefont {Maruyama}}, \bibinfo {author} {\bibfnamefont {H.}~\bibnamefont {Tanaka}}, \bibinfo {author} {\bibfnamefont {W.}~\bibnamefont {Kambara}}, \bibinfo {author} {\bibfnamefont
  {T.}~\bibnamefont {Iwahashi}}, \bibinfo {author} {\bibfnamefont {Y.}~\bibnamefont {Itoh}}, \bibinfo {author} {\bibfnamefont {T.}~\bibnamefont {Osakabe}}, \bibinfo {author} {\bibfnamefont {S.}~\bibnamefont {Wakimoto}}, \bibinfo {author} {\bibfnamefont {K.}~\bibnamefont {Kakurai}}, \bibinfo {author} {\bibfnamefont {F.}~\bibnamefont {Maekawa}}, \bibinfo {author} {\bibfnamefont {M.}~\bibnamefont {Harada}}, \bibinfo {author} {\bibfnamefont {K.}~\bibnamefont {Oikawa}}, \bibinfo {author} {\bibfnamefont {R.}~\bibnamefont {E.~Lechner}}, \bibinfo {author} {\bibfnamefont {F.}~\bibnamefont {Mezei}},\ and\ \bibinfo {author} {\bibfnamefont {M.}~\bibnamefont {Arai}},\ }\bibfield  {title} {\bibinfo {title} {Amateras: A cold-neutron disk chopper spectrometer},\ }\href {https://doi.org/10.1143/JPSJS.80SB.SB028} {\bibfield  {journal} {\bibinfo  {journal} {Journal of the Physical Society of Japan}\ }\textbf {\bibinfo {volume} {80}},\ \bibinfo {pages} {SB028} (\bibinfo {year} {2011})}\BibitemShut {NoStop}%
\bibitem [{\citenamefont {Batyev}\ and\ \citenamefont {Braginski}(1984)}]{Batyev1984}%
  \BibitemOpen
  \bibfield  {author} {\bibinfo {author} {\bibfnamefont {E.~G.}\ \bibnamefont {Batyev}}\ and\ \bibinfo {author} {\bibfnamefont {L.~S.}\ \bibnamefont {Braginski}},\ }\href@noop {} {\bibfield  {journal} {\bibinfo  {journal} {Sov. Phys. JETP}\ }\textbf {\bibinfo {volume} {60}},\ \bibinfo {pages} {781} (\bibinfo {year} {1984})}\BibitemShut {NoStop}%
\bibitem [{\citenamefont {Andreev}\ and\ \citenamefont {Lifshitz}(1971)}]{Andreev1971}%
  \BibitemOpen
  \bibfield  {author} {\bibinfo {author} {\bibfnamefont {A.~F.}\ \bibnamefont {Andreev}}\ and\ \bibinfo {author} {\bibfnamefont {I.~M.}\ \bibnamefont {Lifshitz}},\ }\bibfield  {title} {\bibinfo {title} {Quantum theory of defects in crystals},\ }\href {https://doi.org/10.1070/PU1971v013n05ABEH004235} {\bibfield  {journal} {\bibinfo  {journal} {Soviet Physics Uspekhi}\ }\textbf {\bibinfo {volume} {13}},\ \bibinfo {pages} {670} (\bibinfo {year} {1971})}\BibitemShut {NoStop}%
\bibitem [{\citenamefont {Starykh}(2015)}]{Starykh2015}%
  \BibitemOpen
  \bibfield  {author} {\bibinfo {author} {\bibfnamefont {O.~A.}\ \bibnamefont {Starykh}},\ }\bibfield  {title} {\bibinfo {title} {{Unusual ordered phases of highly frustrated magnets: a review}},\ }\href {https://doi.org/10.1088/0034-4885/78/5/052502} {\bibfield  {journal} {\bibinfo  {journal} {Rep. Prog. Phys.}\ }\textbf {\bibinfo {volume} {78}},\ \bibinfo {pages} {052502} (\bibinfo {year} {2015})}\BibitemShut {NoStop}%
\bibitem [{\citenamefont {Stone}\ \emph {et~al.}(2003)\citenamefont {Stone}, \citenamefont {Reich}, \citenamefont {Broholm}, \citenamefont {Lefmann}, \citenamefont {Rischel}, \citenamefont {Landee},\ and\ \citenamefont {Turnbull}}]{Stone2003}%
  \BibitemOpen
  \bibfield  {author} {\bibinfo {author} {\bibfnamefont {M.~B.}\ \bibnamefont {Stone}}, \bibinfo {author} {\bibfnamefont {D.~H.}\ \bibnamefont {Reich}}, \bibinfo {author} {\bibfnamefont {C.}~\bibnamefont {Broholm}}, \bibinfo {author} {\bibfnamefont {K.}~\bibnamefont {Lefmann}}, \bibinfo {author} {\bibfnamefont {C.}~\bibnamefont {Rischel}}, \bibinfo {author} {\bibfnamefont {C.~P.}\ \bibnamefont {Landee}},\ and\ \bibinfo {author} {\bibfnamefont {M.~M.}\ \bibnamefont {Turnbull}},\ }\bibfield  {title} {\bibinfo {title} {{Extended Quantum Critical Phase in a Magnetized Spin-$1/2$ Antiferromagnetic Chain}},\ }\href {https://doi.org/10.1103/PhysRevLett.91.037205} {\bibfield  {journal} {\bibinfo  {journal} {Phys. Rev. Lett.}\ }\textbf {\bibinfo {volume} {91}},\ \bibinfo {pages} {037205} (\bibinfo {year} {2003})}\BibitemShut {NoStop}%
\bibitem [{\citenamefont {Toth}\ and\ \citenamefont {Lake}(2015)}]{TothLake_JPCM_2015_SpinW}%
  \BibitemOpen
  \bibfield  {author} {\bibinfo {author} {\bibfnamefont {S.}~\bibnamefont {Toth}}\ and\ \bibinfo {author} {\bibfnamefont {B.}~\bibnamefont {Lake}},\ }\bibfield  {title} {\bibinfo {title} {{Linear spin wave theory for single-$Q$ incommensurate magnetic structures}},\ }\href {https://doi.org/10.1088/0953-8984/27/16/166002} {\bibfield  {journal} {\bibinfo  {journal} {J. Phys.: Condens. Matter}\ }\textbf {\bibinfo {volume} {27}},\ \bibinfo {pages} {166002} (\bibinfo {year} {2015})}\BibitemShut {NoStop}%
\bibitem [{\citenamefont {M\"uller}\ \emph {et~al.}(1981)\citenamefont {M\"uller}, \citenamefont {Thomas}, \citenamefont {Beck},\ and\ \citenamefont {Bonner}}]{Muller1981}%
  \BibitemOpen
  \bibfield  {author} {\bibinfo {author} {\bibfnamefont {G.}~\bibnamefont {M\"uller}}, \bibinfo {author} {\bibfnamefont {H.}~\bibnamefont {Thomas}}, \bibinfo {author} {\bibfnamefont {H.}~\bibnamefont {Beck}},\ and\ \bibinfo {author} {\bibfnamefont {J.~C.}\ \bibnamefont {Bonner}},\ }\bibfield  {title} {\bibinfo {title} {Quantum spin dynamics of the antiferromagnetic linear chain in zero and nonzero magnetic field},\ }\href {https://doi.org/10.1103/PhysRevB.24.1429} {\bibfield  {journal} {\bibinfo  {journal} {Phys. Rev. B}\ }\textbf {\bibinfo {volume} {24}},\ \bibinfo {pages} {1429} (\bibinfo {year} {1981})}\BibitemShut {NoStop}%
\bibitem [{\citenamefont {Ulaga}\ \emph {et~al.}(2024)\citenamefont {Ulaga}, \citenamefont {Kokalj}, \citenamefont {Wietek}, \citenamefont {Zorko},\ and\ \citenamefont {Prelov\ifmmode~\check{s}\else \v{s}\fi{}ek}}]{Ulaga2024}%
  \BibitemOpen
  \bibfield  {author} {\bibinfo {author} {\bibfnamefont {M.}~\bibnamefont {Ulaga}}, \bibinfo {author} {\bibfnamefont {J.}~\bibnamefont {Kokalj}}, \bibinfo {author} {\bibfnamefont {A.}~\bibnamefont {Wietek}}, \bibinfo {author} {\bibfnamefont {A.}~\bibnamefont {Zorko}},\ and\ \bibinfo {author} {\bibfnamefont {P.}~\bibnamefont {Prelov\ifmmode~\check{s}\else \v{s}\fi{}ek}},\ }\bibfield  {title} {\bibinfo {title} {Finite-temperature properties of the easy-axis {Heisenberg} model on frustrated lattices},\ }\href {https://doi.org/10.1103/PhysRevB.109.035110} {\bibfield  {journal} {\bibinfo  {journal} {Phys. Rev. B}\ }\textbf {\bibinfo {volume} {109}},\ \bibinfo {pages} {035110} (\bibinfo {year} {2024})}\BibitemShut {NoStop}%
\bibitem [{\citenamefont {Xu}\ \emph {et~al.}(2024)\citenamefont {Xu}, \citenamefont {Hasik}, \citenamefont {Ponsioen},\ and\ \citenamefont {Nevidomskyy}}]{Xu2024}%
  \BibitemOpen
  \bibfield  {author} {\bibinfo {author} {\bibfnamefont {Y.}~\bibnamefont {Xu}}, \bibinfo {author} {\bibfnamefont {J.}~\bibnamefont {Hasik}}, \bibinfo {author} {\bibfnamefont {B.}~\bibnamefont {Ponsioen}},\ and\ \bibinfo {author} {\bibfnamefont {A.~H.}\ \bibnamefont {Nevidomskyy}},\ }\href {https://arxiv.org/abs/2405.05151} {\bibinfo {title} {Simulating spin dynamics of supersolid states in a quantum ising magnet}} (\bibinfo {year} {2024}),\ \Eprint {https://arxiv.org/abs/2405.05151} {arXiv:2405.05151 [cond-mat.str-el]} \BibitemShut {NoStop}%
\bibitem [{\citenamefont {Zhitomirsky}\ and\ \citenamefont {Chernyshev}(2013)}]{ZhitomirskyChernyshev_RMP_2013_DecayReview}%
  \BibitemOpen
  \bibfield  {author} {\bibinfo {author} {\bibfnamefont {M.~E.}\ \bibnamefont {Zhitomirsky}}\ and\ \bibinfo {author} {\bibfnamefont {A.~L.}\ \bibnamefont {Chernyshev}},\ }\bibfield  {title} {\bibinfo {title} {Colloquium: Spontaneous magnon decays},\ }\href {https://doi.org/10.1103/RevModPhys.85.219} {\bibfield  {journal} {\bibinfo  {journal} {Rev. Mod. Phys.}\ }\textbf {\bibinfo {volume} {85}},\ \bibinfo {pages} {219} (\bibinfo {year} {2013})}\BibitemShut {NoStop}%
\bibitem [{\citenamefont {Shu}\ \emph {et~al.}(2023)\citenamefont {Shu}, \citenamefont {Dong}, \citenamefont {Jiao}, \citenamefont {Wu}, \citenamefont {Lin}, \citenamefont {Kamiya}, \citenamefont {Hong}, \citenamefont {Cao}, \citenamefont {Matsuda}, \citenamefont {Tian}, \citenamefont {Chi}, \citenamefont {Ehlers}, \citenamefont {Ouyang}, \citenamefont {Chen}, \citenamefont {Zou}, \citenamefont {Qu}, \citenamefont {Huang}, \citenamefont {Zhou},\ and\ \citenamefont {Ma}}]{Mingfang2023}%
  \BibitemOpen
  \bibfield  {author} {\bibinfo {author} {\bibfnamefont {M.}~\bibnamefont {Shu}}, \bibinfo {author} {\bibfnamefont {W.}~\bibnamefont {Dong}}, \bibinfo {author} {\bibfnamefont {J.}~\bibnamefont {Jiao}}, \bibinfo {author} {\bibfnamefont {J.}~\bibnamefont {Wu}}, \bibinfo {author} {\bibfnamefont {G.}~\bibnamefont {Lin}}, \bibinfo {author} {\bibfnamefont {Y.}~\bibnamefont {Kamiya}}, \bibinfo {author} {\bibfnamefont {T.}~\bibnamefont {Hong}}, \bibinfo {author} {\bibfnamefont {H.}~\bibnamefont {Cao}}, \bibinfo {author} {\bibfnamefont {M.}~\bibnamefont {Matsuda}}, \bibinfo {author} {\bibfnamefont {W.}~\bibnamefont {Tian}}, \bibinfo {author} {\bibfnamefont {S.}~\bibnamefont {Chi}}, \bibinfo {author} {\bibfnamefont {G.}~\bibnamefont {Ehlers}}, \bibinfo {author} {\bibfnamefont {Z.}~\bibnamefont {Ouyang}}, \bibinfo {author} {\bibfnamefont {H.}~\bibnamefont {Chen}}, \bibinfo {author} {\bibfnamefont {Y.}~\bibnamefont {Zou}}, \bibinfo {author} {\bibfnamefont {Z.}~\bibnamefont {Qu}}, \bibinfo {author} {\bibfnamefont
  {Q.}~\bibnamefont {Huang}}, \bibinfo {author} {\bibfnamefont {H.}~\bibnamefont {Zhou}},\ and\ \bibinfo {author} {\bibfnamefont {J.}~\bibnamefont {Ma}},\ }\bibfield  {title} {\bibinfo {title} {Static and dynamical properties of the spin-$\frac{5}{2}$ nearly ideal triangular lattice antiferromagnet {${\mathrm{Ba}}_{3}{\mathrm{MnSb}}_{2}{\mathrm{O}}_{9}$}},\ }\href {https://doi.org/10.1103/PhysRevB.108.174424} {\bibfield  {journal} {\bibinfo  {journal} {Phys. Rev. B}\ }\textbf {\bibinfo {volume} {108}},\ \bibinfo {pages} {174424} (\bibinfo {year} {2023})}\BibitemShut {NoStop}%
\bibitem [{\citenamefont {Sheng}\ \emph {et~al.}(2024)\citenamefont {Sheng}, \citenamefont {Wang}, \citenamefont {Jiang}, \citenamefont {Ge}, \citenamefont {Zhao}, \citenamefont {Li}, \citenamefont {Kofu}, \citenamefont {Yu}, \citenamefont {Zhu}, \citenamefont {Mei}, \citenamefont {Wang},\ and\ \citenamefont {Wu}}]{sheng2024}%
  \BibitemOpen
  \bibfield  {author} {\bibinfo {author} {\bibfnamefont {J.}~\bibnamefont {Sheng}}, \bibinfo {author} {\bibfnamefont {L.}~\bibnamefont {Wang}}, \bibinfo {author} {\bibfnamefont {W.}~\bibnamefont {Jiang}}, \bibinfo {author} {\bibfnamefont {H.}~\bibnamefont {Ge}}, \bibinfo {author} {\bibfnamefont {N.}~\bibnamefont {Zhao}}, \bibinfo {author} {\bibfnamefont {T.}~\bibnamefont {Li}}, \bibinfo {author} {\bibfnamefont {M.}~\bibnamefont {Kofu}}, \bibinfo {author} {\bibfnamefont {D.}~\bibnamefont {Yu}}, \bibinfo {author} {\bibfnamefont {W.}~\bibnamefont {Zhu}}, \bibinfo {author} {\bibfnamefont {J.-W.}\ \bibnamefont {Mei}}, \bibinfo {author} {\bibfnamefont {Z.}~\bibnamefont {Wang}},\ and\ \bibinfo {author} {\bibfnamefont {L.}~\bibnamefont {Wu}},\ }\href@noop {} {\bibinfo {title} {Continuum of spin excitations in an ordered magnet}} (\bibinfo {year} {2024}),\ \Eprint {https://arxiv.org/abs/2402.07730} {arXiv:2402.07730 [cond-mat.str-el]} \BibitemShut {NoStop}%
\bibitem [{\citenamefont {Wessel}\ and\ \citenamefont {Troyer}(2005)}]{Wessel2005}%
  \BibitemOpen
  \bibfield  {author} {\bibinfo {author} {\bibfnamefont {S.}~\bibnamefont {Wessel}}\ and\ \bibinfo {author} {\bibfnamefont {M.}~\bibnamefont {Troyer}},\ }\bibfield  {title} {\bibinfo {title} {Supersolid hard-core bosons on the triangular lattice},\ }\href {https://doi.org/10.1103/PhysRevLett.95.127205} {\bibfield  {journal} {\bibinfo  {journal} {Phys. Rev. Lett.}\ }\textbf {\bibinfo {volume} {95}},\ \bibinfo {pages} {127205} (\bibinfo {year} {2005})}\BibitemShut {NoStop}%
\bibitem [{\citenamefont {Jia}\ \emph {et~al.}(2023)\citenamefont {Jia}, \citenamefont {Ma}, \citenamefont {Wang},\ and\ \citenamefont {Chen}}]{jia2023}%
  \BibitemOpen
  \bibfield  {author} {\bibinfo {author} {\bibfnamefont {H.}~\bibnamefont {Jia}}, \bibinfo {author} {\bibfnamefont {B.}~\bibnamefont {Ma}}, \bibinfo {author} {\bibfnamefont {Z.}~\bibnamefont {Wang}},\ and\ \bibinfo {author} {\bibfnamefont {G.}~\bibnamefont {Chen}},\ }\href@noop {} {\bibinfo {title} {Quantum spin supersolid as a precursory dirac spin liquid in a triangular lattice antiferromagnet}} (\bibinfo {year} {2023}),\ \Eprint {https://arxiv.org/abs/2304.11716} {arXiv:2304.11716 [cond-mat.str-el]} \BibitemShut {NoStop}%
\bibitem [{\citenamefont {Ghasemi}\ \emph {et~al.}()\citenamefont {Ghasemi}, \citenamefont {Chen}, \citenamefont {Zhang}, \citenamefont {Shi}, \citenamefont {Choi}, \citenamefont {Lee}, \citenamefont {Hao}, \citenamefont {Cao}, \citenamefont {Winn}, \citenamefont {Jaime}, \citenamefont {Xu}, \citenamefont {Zhong}, \citenamefont {Cava}, \citenamefont {Armitage},\ and\ \citenamefont {Broholm}}]{Collin}%
  \BibitemOpen
  \bibfield  {author} {\bibinfo {author} {\bibfnamefont {A.}~\bibnamefont {Ghasemi}}, \bibinfo {author} {\bibfnamefont {T.}~\bibnamefont {Chen}}, \bibinfo {author} {\bibfnamefont {J.}~\bibnamefont {Zhang}}, \bibinfo {author} {\bibfnamefont {L.}~\bibnamefont {Shi}}, \bibinfo {author} {\bibfnamefont {E.-S.}\ \bibnamefont {Choi}}, \bibinfo {author} {\bibfnamefont {M.}~\bibnamefont {Lee}}, \bibinfo {author} {\bibfnamefont {Y.}~\bibnamefont {Hao}}, \bibinfo {author} {\bibfnamefont {H.}~\bibnamefont {Cao}}, \bibinfo {author} {\bibfnamefont {B.}~\bibnamefont {Winn}}, \bibinfo {author} {\bibfnamefont {M.}~\bibnamefont {Jaime}}, \bibinfo {author} {\bibfnamefont {X.}~\bibnamefont {Xu}}, \bibinfo {author} {\bibfnamefont {R.}~\bibnamefont {Zhong}}, \bibinfo {author} {\bibfnamefont {R.}~\bibnamefont {Cava}}, \bibinfo {author} {\bibfnamefont {P.}~\bibnamefont {Armitage}},\ and\ \bibinfo {author} {\bibfnamefont {C.}~\bibnamefont {Broholm}},\ }\href@noop {} {}\bibinfo {note} {Quantum fluctuations in an easy axis spin-1/2
  triangular lattice antiferromagnet. APS March Meeting 2024, Abstract no. Y23.00007}\BibitemShut {NoStop}%
\end{thebibliography}%


\begin{thebibliography}{2}%
\makeatletter
\providecommand \@ifxundefined [1]{%
 \@ifx{#1\undefined}
}%
\providecommand \@ifnum [1]{%
 \ifnum #1\expandafter \@firstoftwo
 \else \expandafter \@secondoftwo
 \fi
}%
\providecommand \@ifx [1]{%
 \ifx #1\expandafter \@firstoftwo
 \else \expandafter \@secondoftwo
 \fi
}%
\providecommand \natexlab [1]{#1}%
\providecommand \enquote  [1]{``#1''}%
\providecommand \bibnamefont  [1]{#1}%
\providecommand \bibfnamefont [1]{#1}%
\providecommand \citenamefont [1]{#1}%
\providecommand \href@noop [0]{\@secondoftwo}%
\providecommand \href [0]{\begingroup \@sanitize@url \@href}%
\providecommand \@href[1]{\@@startlink{#1}\@@href}%
\providecommand \@@href[1]{\endgroup#1\@@endlink}%
\providecommand \@sanitize@url [0]{\catcode `\\12\catcode `\$12\catcode `\&12\catcode `\#12\catcode `\^12\catcode `\_12\catcode `\%12\relax}%
\providecommand \@@startlink[1]{}%
\providecommand \@@endlink[0]{}%
\providecommand \url  [0]{\begingroup\@sanitize@url \@url }%
\providecommand \@url [1]{\endgroup\@href {#1}{\urlprefix }}%
\providecommand \urlprefix  [0]{URL }%
\providecommand \Eprint [0]{\href }%
\providecommand \doibase [0]{https://doi.org/}%
\providecommand \selectlanguage [0]{\@gobble}%
\providecommand \bibinfo  [0]{\@secondoftwo}%
\providecommand \bibfield  [0]{\@secondoftwo}%
\providecommand \translation [1]{[#1]}%
\providecommand \BibitemOpen [0]{}%
\providecommand \bibitemStop [0]{}%
\providecommand \bibitemNoStop [0]{.\EOS\space}%
\providecommand \EOS [0]{\spacefactor3000\relax}%
\providecommand \BibitemShut  [1]{\csname bibitem#1\endcsname}%
\let\auto@bib@innerbib\@empty
\bibitem [{\citenamefont {Zavareh}\ \emph {et~al.}(2017)\citenamefont {Zavareh}, \citenamefont {Skourski}, \citenamefont {Skokov}, \citenamefont {Karpenkov}, \citenamefont {Zvyagina}, \citenamefont {Waske}, \citenamefont {Haskel}, \citenamefont {Zhernenkov}, \citenamefont {Wosnitza},\ and\ \citenamefont {Gutfleisch}}]{Zavareh2017}%
  \BibitemOpen
  \bibfield  {author} {\bibinfo {author} {\bibfnamefont {M.~G.}\ \bibnamefont {Zavareh}}, \bibinfo {author} {\bibfnamefont {Y.}~\bibnamefont {Skourski}}, \bibinfo {author} {\bibfnamefont {K.~P.}\ \bibnamefont {Skokov}}, \bibinfo {author} {\bibfnamefont {D.~Y.}\ \bibnamefont {Karpenkov}}, \bibinfo {author} {\bibfnamefont {L.}~\bibnamefont {Zvyagina}}, \bibinfo {author} {\bibfnamefont {A.}~\bibnamefont {Waske}}, \bibinfo {author} {\bibfnamefont {D.}~\bibnamefont {Haskel}}, \bibinfo {author} {\bibfnamefont {M.}~\bibnamefont {Zhernenkov}}, \bibinfo {author} {\bibfnamefont {J.}~\bibnamefont {Wosnitza}},\ and\ \bibinfo {author} {\bibfnamefont {O.}~\bibnamefont {Gutfleisch}},\ }\bibfield  {title} {\bibinfo {title} {Direct measurement of the magnetocaloric effect in $\mathrm{La}(\mathrm{Fe},\mathrm{Si},\mathrm{Co}{)}_{13}$ compounds in pulsed magnetic fields},\ }\href {https://doi.org/10.1103/PhysRevApplied.8.014037} {\bibfield  {journal} {\bibinfo  {journal} {Phys. Rev. Appl.}\ }\textbf {\bibinfo {volume} {8}},\
  \bibinfo {pages} {014037} (\bibinfo {year} {2017})}\BibitemShut {NoStop}%
\bibitem [{\citenamefont {Kamiya}\ \emph {et~al.}(2018)\citenamefont {Kamiya}, \citenamefont {Ge}, \citenamefont {Hong}, \citenamefont {Qiu}, \citenamefont {Quintero-Castro}, \citenamefont {Lu}, \citenamefont {Cao}, \citenamefont {Matsuda}, \citenamefont {Choi}, \citenamefont {Batista}, \citenamefont {Mourigal}, \citenamefont {D.},\ and\ \citenamefont {Ma}}]{KamiyaGe_NatComm_2018_Ba3CoSb2O9spectrumUUD}%
  \BibitemOpen
  \bibfield  {author} {\bibinfo {author} {\bibfnamefont {Y.}~\bibnamefont {Kamiya}}, \bibinfo {author} {\bibfnamefont {L.}~\bibnamefont {Ge}}, \bibinfo {author} {\bibfnamefont {T.}~\bibnamefont {Hong}}, \bibinfo {author} {\bibfnamefont {Y.}~\bibnamefont {Qiu}}, \bibinfo {author} {\bibfnamefont {D.~L.}\ \bibnamefont {Quintero-Castro}}, \bibinfo {author} {\bibfnamefont {Z.}~\bibnamefont {Lu}}, \bibinfo {author} {\bibfnamefont {H.~B.}\ \bibnamefont {Cao}}, \bibinfo {author} {\bibfnamefont {M.}~\bibnamefont {Matsuda}}, \bibinfo {author} {\bibfnamefont {E.~S.}\ \bibnamefont {Choi}}, \bibinfo {author} {\bibfnamefont {C.~D.}\ \bibnamefont {Batista}}, \bibinfo {author} {\bibfnamefont {M.}~\bibnamefont {Mourigal}}, \bibinfo {author} {\bibfnamefont {Z.~H.}\ \bibnamefont {D.}},\ and\ \bibinfo {author} {\bibfnamefont {J.}~\bibnamefont {Ma}},\ }\bibfield  {title} {\bibinfo {title} {{The nature of spin excitations in the one-third magnetization plateau phase of Ba$_3$CoSb$_2$O$_9$}},\ }\href
  {https://doi.org/10.1038/s41467-018-04914-1} {\bibfield  {journal} {\bibinfo  {journal} {Nat. Commun.}\ }\textbf {\bibinfo {volume} {9}},\ \bibinfo {pages} {2666} (\bibinfo {year} {2018})}\BibitemShut {NoStop}%
\end{thebibliography}%
\end{document}


\title{Continuum excitations in a spin-supersolid on a triangular lattice:\\
Supplementary material}

\author{M.~Zhu}
\address{Laboratory for Solid State Physics, ETH Z\"{u}rich, 8093 Z\"{u}rich, Switzerland}
\author{V. Romerio}
\address{Laboratory for Solid State Physics, ETH Z\"{u}rich, 8093 Z\"{u}rich, Switzerland}
\author{N. Steiger}
\address{Laboratory for Solid State Physics, ETH Z\"{u}rich, 8093 Z\"{u}rich, Switzerland}
\author{S. D. Nabi}
\address{Laboratory for Solid State Physics, ETH Z\"{u}rich, 8093 Z\"{u}rich, Switzerland}
\author{N. Murai}
\address{J-PARC Center, Japan Atomic Energy Agency, Tokai, Ibaraki 319-1195, Japan}

\author{S. Ohira-Kawamura}	
\address{J-PARC Center, Japan Atomic Energy Agency, Tokai, Ibaraki 319-1195, Japan}

\author{K.~Yu.~Povarov}
\address{Dresden High Magnetic Field Laboratory (HLD-EMFL) and W\"urzburg-Dresden Cluster of Excellence ct.qmat, Helmholtz-Zentrum Dresden-Rossendorf, 01328 Dresden, Germany}
\author{Y. Skourski}
\address{Dresden High Magnetic Field Laboratory (HLD-EMFL) and W\"urzburg-Dresden Cluster of Excellence ct.qmat, Helmholtz-Zentrum Dresden-Rossendorf, 01328 Dresden, Germany}

\author{R. Sibille}
\address{Laboratory for Neutron Scattering and Imaging, Paul Scherrer Institute, CH-5232 Villigen, Switzerland}
\author{L. Keller}
\address{Laboratory for Neutron Scattering and Imaging, Paul Scherrer Institute, CH-5232 Villigen, Switzerland}

\author{Z. Yan}
\address{Laboratory for Solid State Physics, ETH Z\"{u}rich, 8093 Z\"{u}rich, Switzerland}
\author{S. Gvasaliya}
\address{Laboratory for Solid State Physics, ETH Z\"{u}rich, 8093 Z\"{u}rich, Switzerland}

\author{A.~Zheludev}
\email{zhelud@ethz.ch; http://www.neutron.ethz.ch/}
\address{Laboratory for Solid State Physics, ETH Z\"{u}rich, 8093 Z\"{u}rich, Switzerland}

\date{\today}
\maketitle
\section{Calorimetry data}
The false-color specific heat plot shown in Fig.~2b in the main text combines a series of constant-field temperature scans. Typical low-temperature data are shown in Fig.~\ref{sup_bulk}.

{
\section{Influence of magnetocaloric effect on pulsed-field magnetometry}
The magnetization data (solid lines) shown in Fig.~2a of the main text was collected using a pulsed-field magnetometer, where the duration of each pulse is 0.1 sec. Since the pulse is so short the sample does not have enough time to exchange heat with the thermostat to always reach thermal equilibrium as the field sweeps. The sample temperature will change due to the magnetocaloric effect. As a result, the magnetization curve as well as the value of the critical field, may appear different during field-sweeping up and sweeping-down, and from that measured under isothermal conditions. This is a known drawback of this measurement technique. On the same instrument as used in the present study similar effects have been reported and discussed for example in Ref.~\cite{Zavareh2017}. 

In contrast, the magnetization data from the Faraday balance magnetometry (symbols in Fig.~2a) was obtained under isothermal conditions. The critical field for the transition into the plateau phase agrees very well with that determined by the specific heat data (Fig.~2b). }

{ 
\section{Neutron diffraction}
Fig.~\ref{sup_diff_cut}(a) and (c) presents the individual $l$-scans over the magnetic Bragg peaks that contributed to the averaged curves of Fig.~3b. The ``A'' and ``B'' labels correspond to two families of equivalent reflections, as labeled in Fig.~3a. {All such scans were obtained by integrating the measured intensity in the range $\pm0.02$ r.l.u. along the (1,1,0) and (1,-1,0) directions of the reciprocal space and $\pm0.2$~meV in energy.} 
Although the 2-dimensional Bragg positions are equivalent in each set, the corresponding scans show considerable variation and some asymmetry. One reason for this is imperfect background subtraction. This may be particularly important for ``A'' type reflections that are positioned on top of some powder diffraction line coming from sample environment. The asymmetry is most visible for the smaller-angle ``B'' reflections. It is likely due to an imperfect vertical centering of the sample within the gap in the split-coil cryomagnet. This results in unequal partial attenuation of the scattered beam at large positive and large negative out-of-plane scattering angles, respectively. At a given value of $l$, the smaller-angle data will be those more strongly affected. In our data analysis we deliberately averaged the scans in each set, and used two symmetric peaks at $l=\pm1/2$, and fixed the Lorentz width of the Voigt functions to be global for ``A" and ``B"reflections in the fit, in hope to partially cancel out these systematic errors.

The same data collected at $\mu_0H=1.5$~T in the plateau phase are shown in panels (b) and (d). Analogs of Fig.~3(a) and (b) corresponding to these conditions are shown in Fig.~\ref{sup_diff}. These data are very similar to what is seen in zero field, except for an increased overall intensity of magnetic scattering. The inter-plane correlation length at $\mu_0H=1.5$~T is determined by the fit described in the main text to be $\lambda \approx 13(1)$ \AA, or about 0.7 lattice unit (solid lines in Fig.~\ref{sup_diff}b). }

Unfortunately, the temperature dependence of magnetic intensity at $\mu_0H=1.5$~T was not measured in our ZEBRA experiment. Instead, Fig.~\ref{sup_diff_t} compares the gradual temperature evolution of the magnetic intensity at $(1/3,1/3,0)$ seen in zero fields (same data as in Fig.~3c in the main text) with clear order-parameter-like behavior deep in the $uud$ phase, at $\mu_0H=6$~T.

{ \section{Inelastic neutron scattering }}

\subsection{ Integration procedure for inelastic data}
{ The inelastic neutron scattering (INS) intensity of Fig.~4(a) and (b) of the main text are obtained by integrating the INS data along the $l$ axis in the entire range, and within $\pm0.1$ r.l.u. along the transverse direction within the (0,0,$1$)-plane. The integration range is the same in the energy-momentum plots in Fig.~\ref{spectra_cut_sup} and \ref{spectra_sup}. Fig.~4(c) and 4(f) are obtained by integrating the INS data along $l$ axis in the entire range and in energy between 0.16-0.18 meV and 0.1-0.9 meV, respectively. The data presented in Fig.~\ref{spectra_sup_consE} are obtained similarly by integrating over different energy intervals as noted in each panel. }

{ 
\subsection{Additional data}
Typical energy-momentum cuts of the inelastic neutron scattering data along high-symmetry directions in the 2D Brillouin zone at zero field and in a magnetic field  $\mu_0H=1.5$~T are presented in Fig.~\ref{spectra_cut_sup}.
Typical constant-energy maps are shown in Fig.~\ref{spectra_sup_consE}. Typical constant-\textbf{Q} cuts are shown in Fig.~\ref{spectra_sup_consQ}. Note the y-axis is plotted in logarithmic scale to highlight the continuum of excitations. The corresponding plots with linear y-axis is shown as insets. The instrumental energy resolution is represented by the shaded Gaussian function. The latter is arbitrarily scaled and provided for reference only. Percentages in (b),(c) and (d) indicated a conservative lower estimate for the fraction of total energy-integrated intensity that can not be attributed to any resolution-limited features, such as a delta function or an algebraic divergence at the lower continuum bound. 
}

\subsection{SWT calculations for $H=0$}
Due to the classical degeneracy in zero applied field, the $Y$-structure is not a unique classical ground state. In order to perform an SWT calculation for $H\rightarrow0$, we assumed a tiny $H=30$~G field along the $c$ axis. This uniquely selects the $Y$ structure as the ground state but opens a gap $\Delta\sim 0.04$~meV in one of the spin wave branches. This artificially-induced residual gap is apparent in all zero-field simulations near the magnetic Bragg points.

\subsection{SWT calculations and fit for $\mu_0H=1.5$~T}
It is instructive to attempt refining the Hamiltonian parameters by fitting the dispersion relation measured in the uud phase at $\mu_0H=1.5$~T using linear SWT. First, we used simplistic Gaussian fits to the data to quantify the measured dispersion relation. The result was then fit to linear SWT to obtain the following set of effective parameters:  $\tilde{g}$ = 8.34(6), $\tilde{J}_{zz}$ = 0.68(6)~meV and $\tilde{J}_{xy}$ = 0.25(1)~meV. Compared to the actual exchange constants, these values are drastically renormalized, particularly the value of $J_{zz}$. The effect is even larger than in other known triangular-lattice systems \cite{KamiyaGe_NatComm_2018_Ba3CoSb2O9spectrumUUD}.
The fitted parameters are also entirely incompatible with magnetization data, underestimating the saturation field by a factor of 4. 

In Fig.~\ref{spectra_sup} we compare the simulated  spectra for a wider energy range up to 1.8 meV using the exchange parameters and gyromagnetic ratio deduced from the magnetization data and quoted in the main text (Fig.~\ref{spectra_sup}a,c) with those computed using the fit result (Fig.~\ref{spectra_sup}b,d).
The fitted parameters do reproduce the curvature of the upper spin wave branch better than the actual values. On the other hand, they predict an additional low-lying spin wave branch that is not observed experimentally. Clearly they are inadequate for describing the physics of the system. This exercise mainly serves to illustrate that, despite being a colinear structure with the same magnetization as predicted by the classical model, the uud state is anything but classical. Any attempt at a quasiclassical description requires a huge renormalization of Hamiltonian parameters.

\renewcommand{\thefigure}{S1}
\begin{figure*}
	\includegraphics[width=\textwidth]{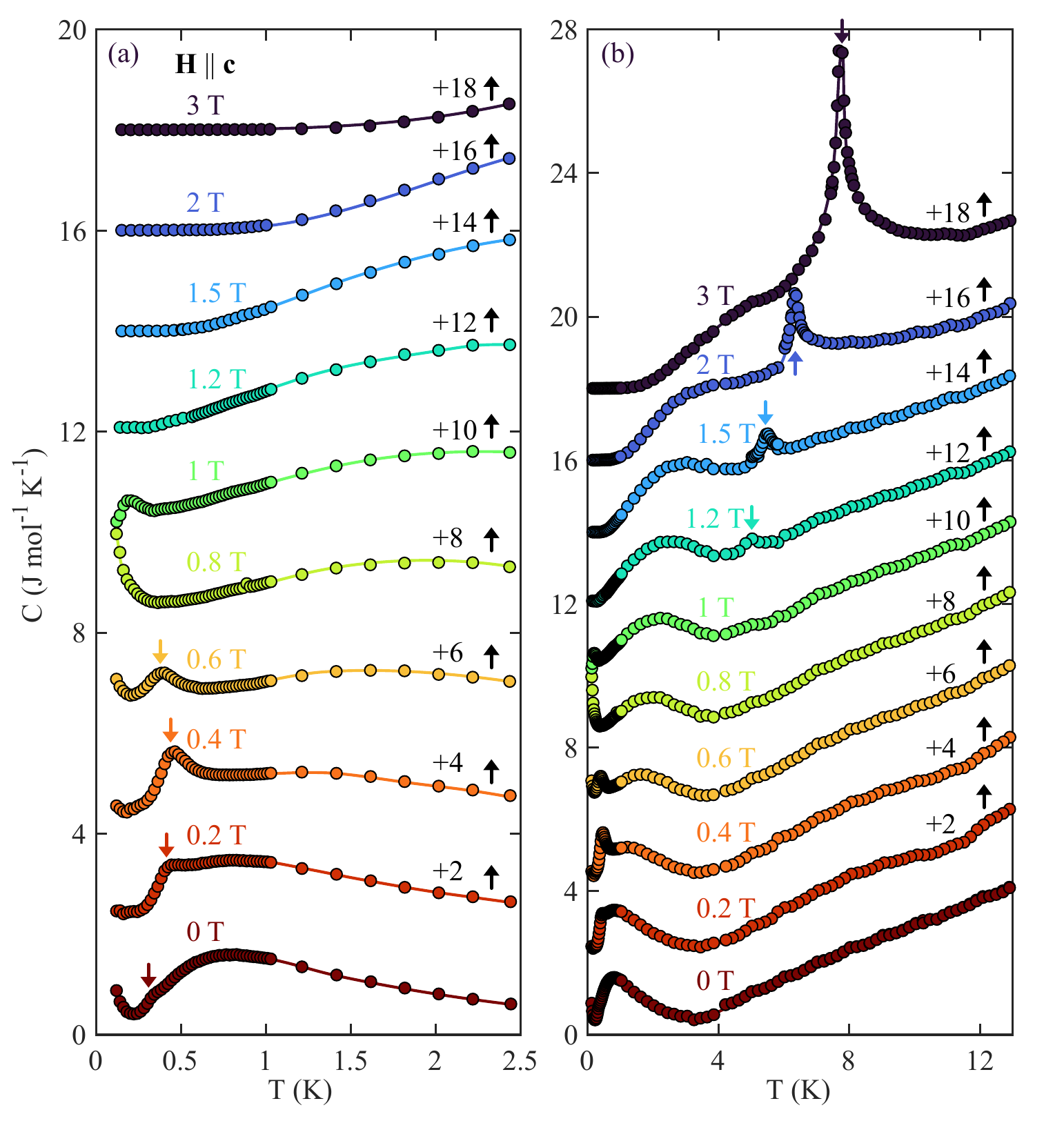}
	\caption{(a) Typical constant-field scans of specific heat, as measured in \KSeCo using the thermal-relaxation method in different magnetic fields applied along the $c$ axis. Individual plots are offset for visibility as indicated on the right. Arrows on the left indicate anomalies on the boundary of the low-temperature spin-supersolid phase. { (b) shows the same data but up to 13 K. The colored arrows denote the $\lambda$-anomaly corresponding to the high-temperature phase boundary of the $uud$ phase (dotted lines in Fig.~2b). } }\label{sup_bulk}
\end{figure*}

\renewcommand{\thefigure}{S2}
\begin{figure*}
	\includegraphics[width=\textwidth]{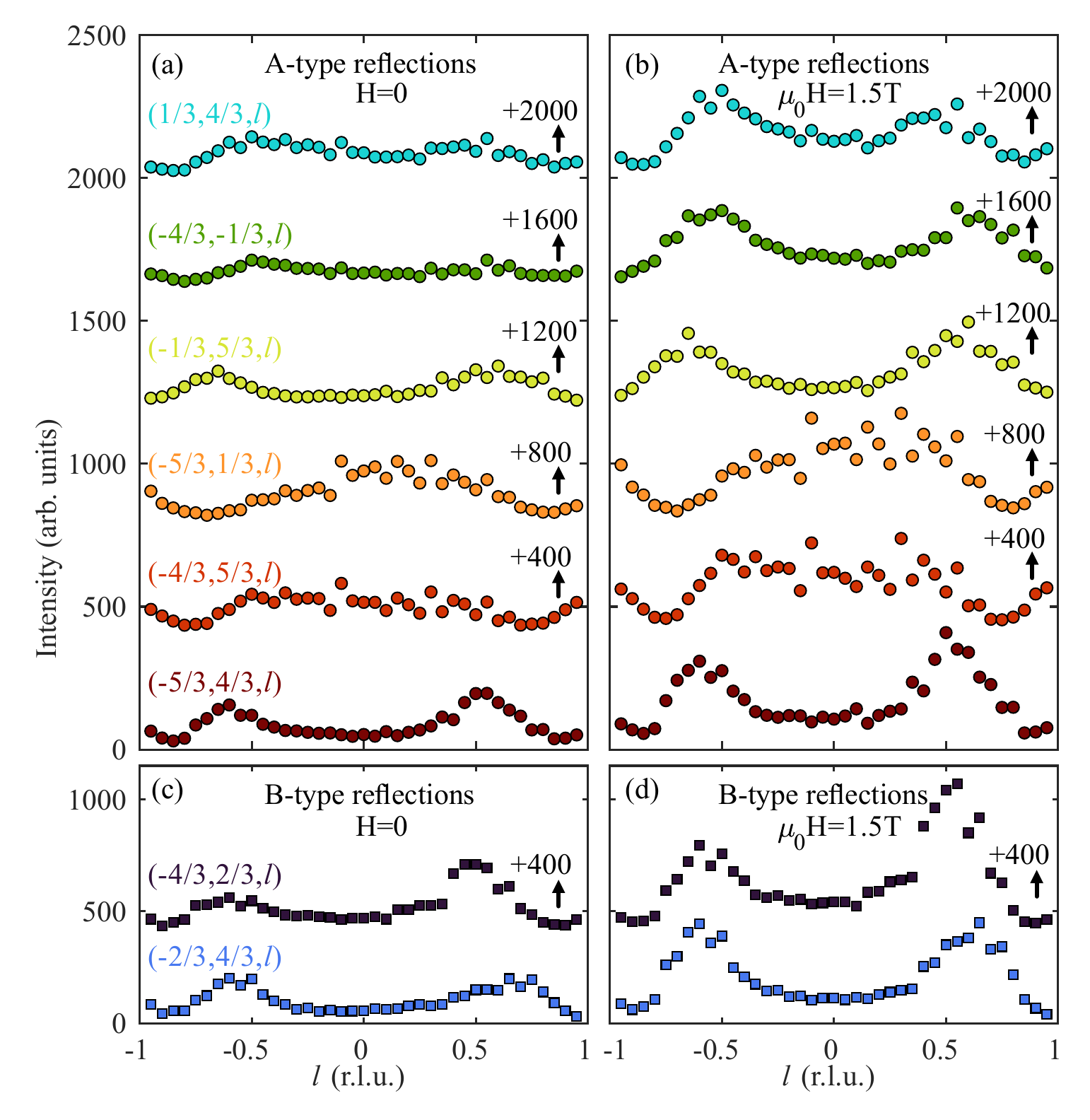}
	\caption{{ Representative $l$-scans of the elastic scattering measured at two sets ``A'' and ``B'' of equivalent  $(1/3,1/3)$-type Bragg positions at $T=70$~mK in \KSeCo (a),(c) are taken in zero field and (c),(d) in a magnetic field $\mu_0H=1.5$~T applied along the $c$ axis. The background measured just off the $(1/3,1/3)$ position has been subtracted point-by-point from each scan. Statistical error bars are smaller than symbol size. Individual cut has been offset for clarity}}\label{sup_diff_cut}
\end{figure*}

\renewcommand{\thefigure}{S3}
\begin{figure*}
	\includegraphics[width=\textwidth]{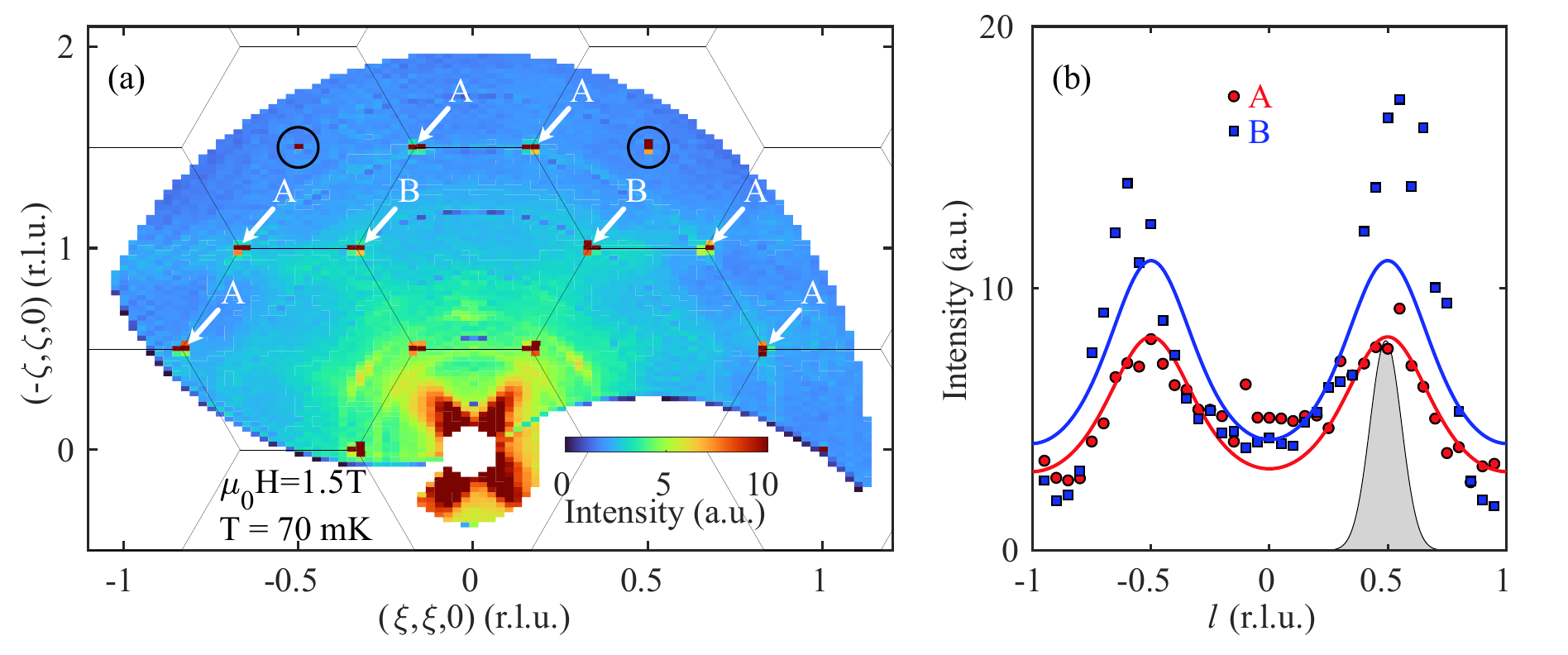}
	\caption{Elastic scattering measured in \KSeCo in a magnetic field $\mu_0H=1.5$~T. (a) TOF data collected  at $T=70$~mK in the $(h,k,0)$ plane reveal $(1/3,1/3)$-type two-dimensional order (arrows). Circles highlight allowed structural Bragg peaks. Solid lines are 2D Brillouin zone boundaries for a single triangular plane. (b) Cuts from the same data set along the $l$ direction { averaged over two equivalent sets of magnetic Bragg rods $A$ and $B$} show only short-range correlations between planes. The shaded Gaussian represent the experimental wave vector resolution { as determined in a fit to the (-2,1,0) nuclear Bragg reflection}. { The background measured just off the $(1/3,1/3)$ position has been subtracted point-by-point. Statistical error bars are smaller than symbol size. The solid lines are fit to the data similar to those described in the main text.} }\label{sup_diff}
\end{figure*}

\renewcommand{\thefigure}{S4}
\begin{figure}
	\includegraphics[width=\columnwidth]{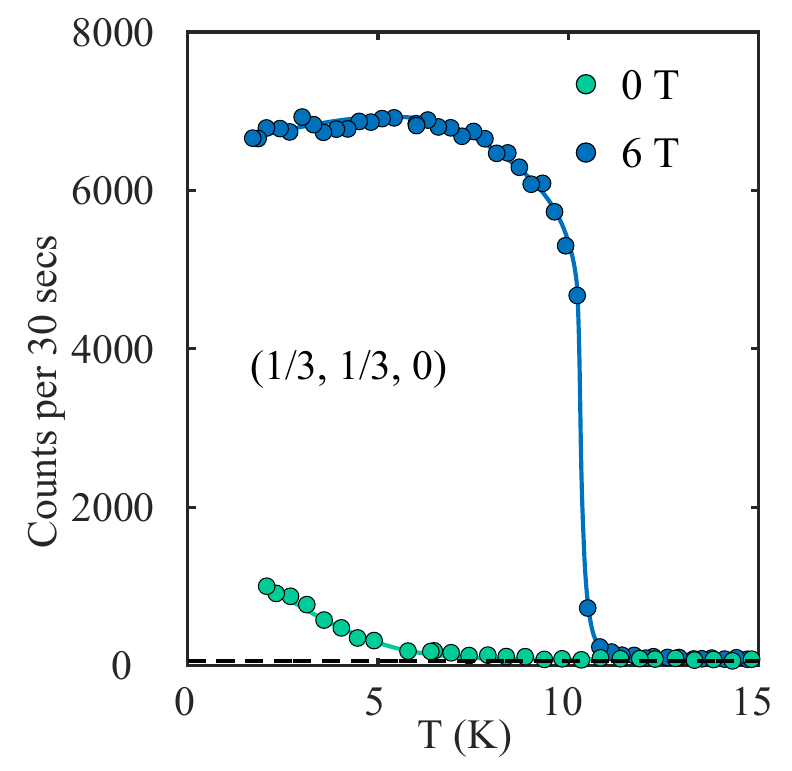}
	\caption{Temperature dependence of neutron diffraction intensity measured at $(1/3,1/3,0)$ in zero field and at $\mu_0H=6$~T. Dashed line is the background measured slightly off the Bragg rod. Error bars are smaller than symbol size.}\label{sup_diff_t}
\end{figure}

\renewcommand{\thefigure}{S5}
\begin{figure*}
	\includegraphics[width=\textwidth]{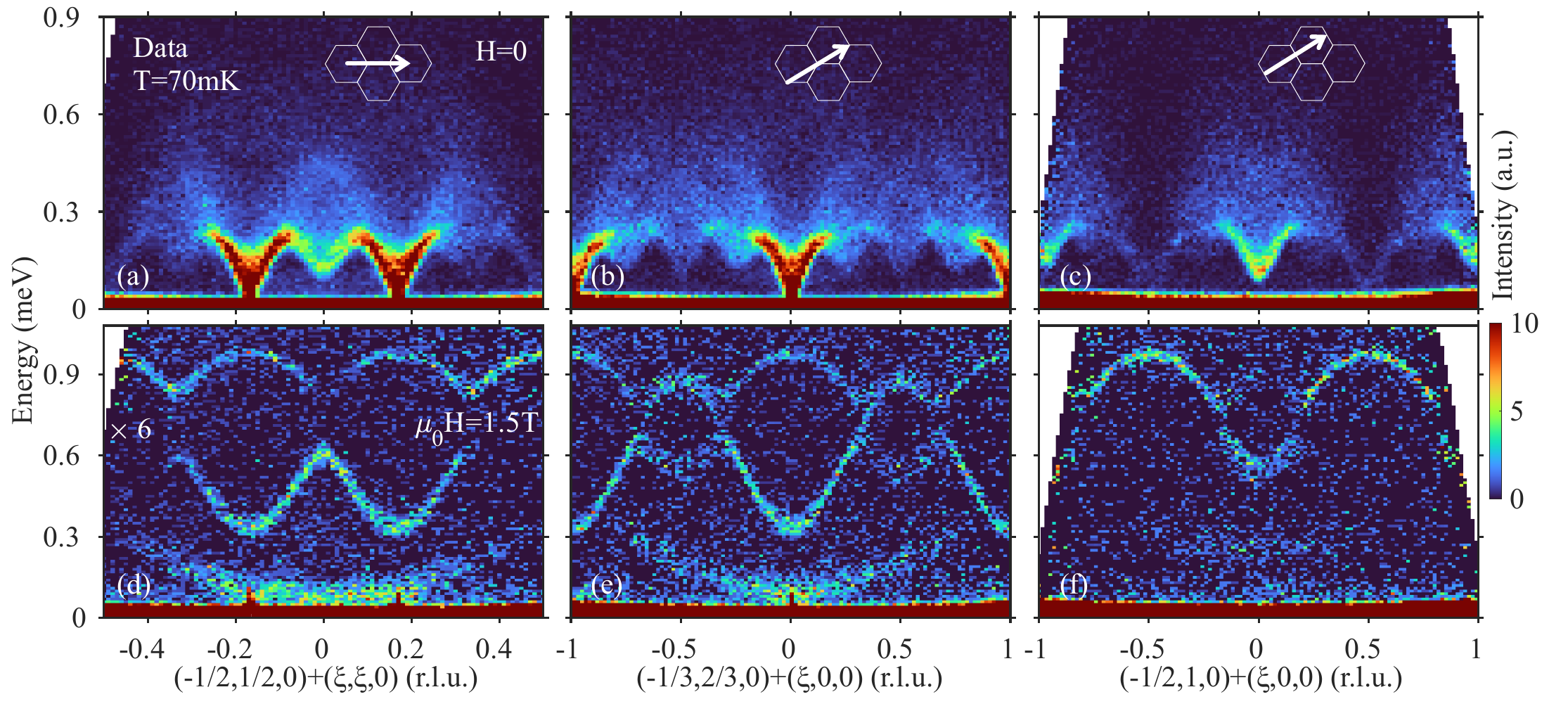}
	\caption{{ Typical false-color plots of the energy-momentum cuts of inelastic neutron intensity measured in \KSeCo at $T=70$~mK in zero field (a-c), and in $\mu_0H=1.5$~T applied along the $c$ axis (d-f). In the latter, note the $\times 6$ scale factor. No background is subtracted. The low-energy scattering below 0.3 meV in (d) and (e) is spurious, originating from the sample environment. It is also present in (a) and (b) but not visible due to a different color range. { Insets show the corresponding paths in the reciprocal space, where hexagons represent the 2D Brillouin zone boundaries. }
		}}\label{spectra_cut_sup}
\end{figure*}

\renewcommand{\thefigure}{S6}
\begin{figure*}
	\includegraphics[width=\textwidth]{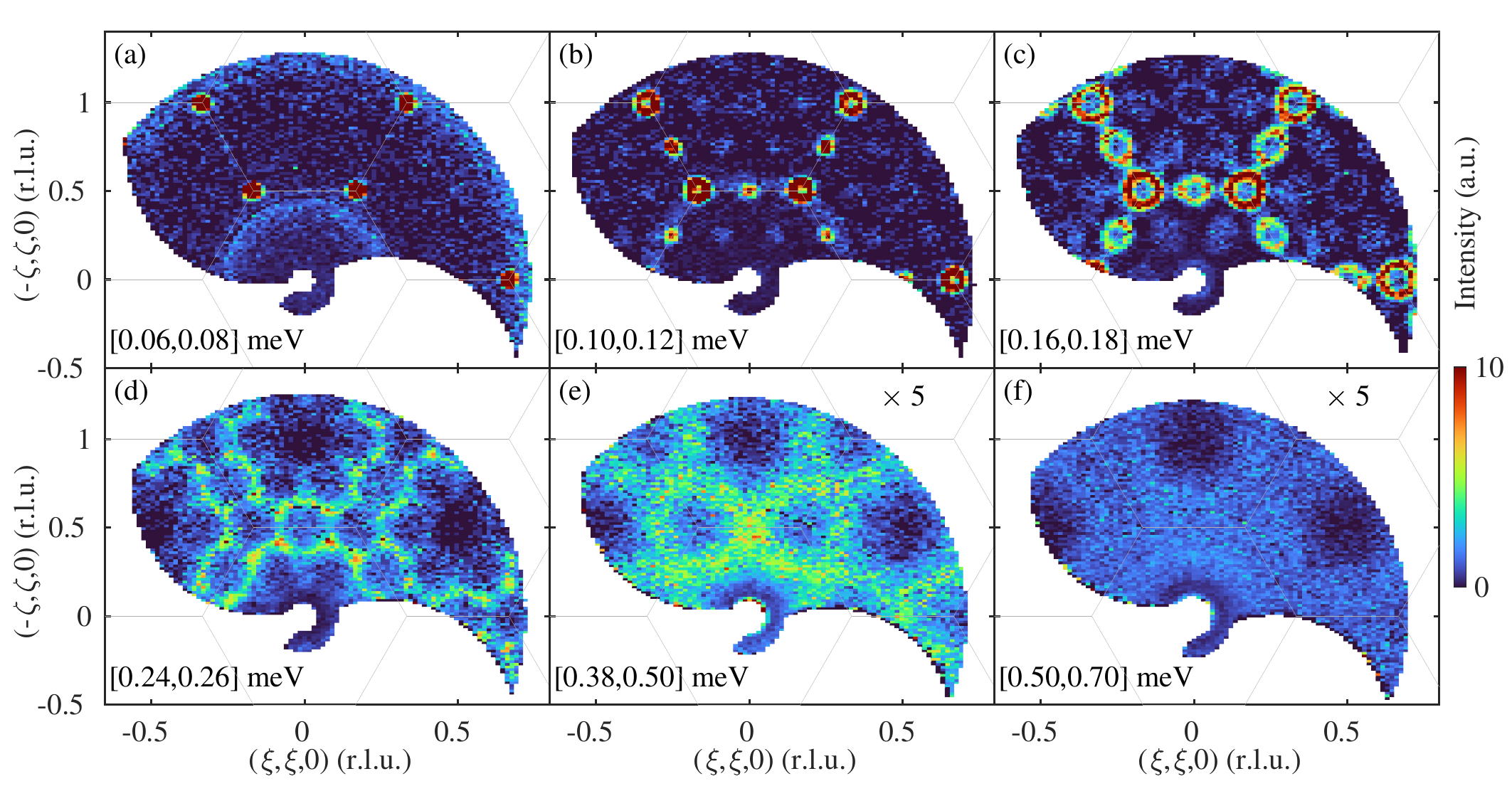}
	\caption{{ Constant-energy maps of inelastic neutron intensity measured in \KSeCo at $T=70$~mK in zero field. Note the $\times 5$ scale factor in (e) and (f). The data were integrated along the $l$ axis in the entire range. No background is subtracted. 
		}} \label{spectra_sup_consE}
\end{figure*}

\renewcommand{\thefigure}{S7}
\begin{figure*}
	\includegraphics[width=0.55\textwidth]{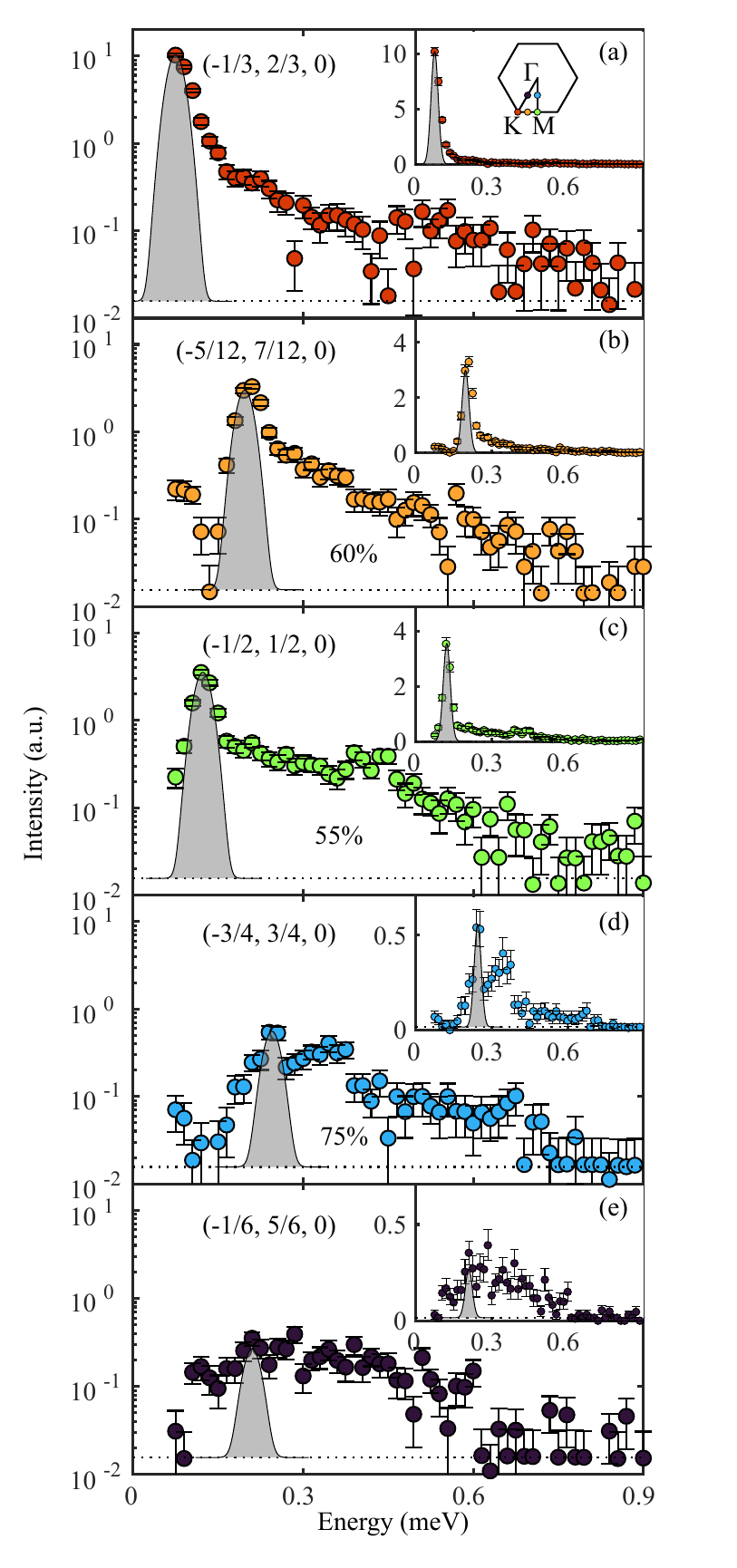}
	\caption{Constant-\textbf{Q} cuts of inelastic neutron intensity measured in \KSeCo at $T=70$~mK in zero field for representative wave vectors.  The data were integrated along the $l$ axis in the entire range and by $\pm$0.03 r.l.u along two orthorgonal directions in the (0,0,1)-plane. Note the y-axis is in logarithmic scale. Insets show the corresponding plots with y-axis in linear scale. The shaded Gaussian represents the instrumental energy resolution. Dashed lines denote the non-magnetic background. Hexagon in (a) illustrates the 2D Brillouin zone boundary and the locations of these wave vectors. Percentages in (b),(c) and (d) indicated a conservative lower estimate for the fraction of total energy-integrated intensity that can not be attributed to any resolution-limited features, such as a delta function or an algebraic divergence at the lower continuum bound.} \label{spectra_sup_consQ}
\end{figure*}

\renewcommand{\thefigure}{S8}
\begin{figure*}
	\includegraphics[width=\textwidth]{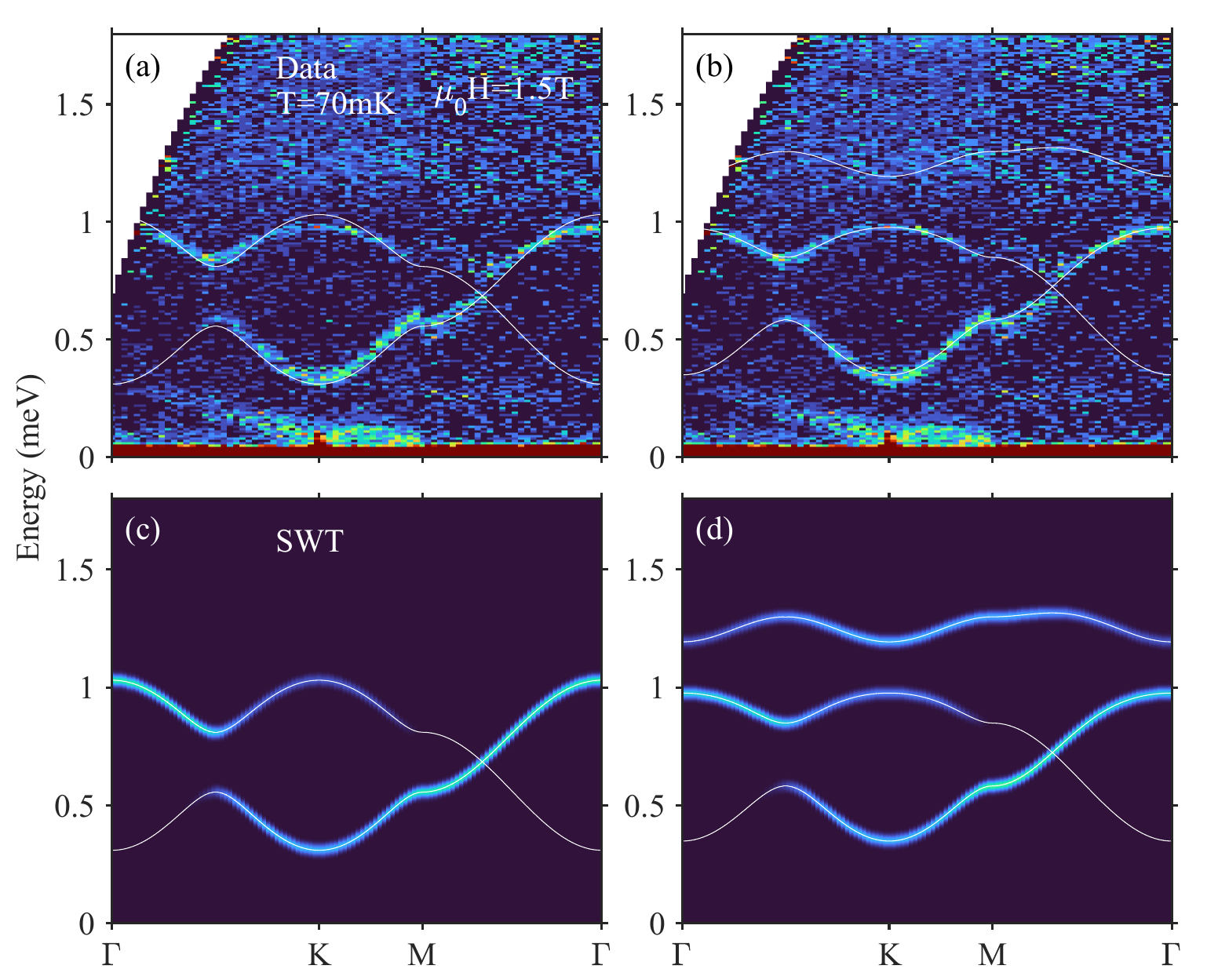}
	\caption{(a),(b) False-color plots of inelastic neutron intensity measured in \KSeCo at $T=70$~mK in $\mu_0H=1.5$~T applied along the $c$ axis. 
    (c),(d) as well as white lines in all panels show corresponding simulations based on linear spin wave theory. (a) and (c) used the real exchange parameters and gyromagnetic ratio deduced from the transition fields in the magnetization data, whereas ``renormalized'' values obtained by fitting the two low-energy branches are used in (b) and (d), respectively, as described in the text.
}\label{spectra_sup}
\end{figure*}

\bibliography{SM_KSeCoSup_19_08_2024}